\documentclass[%
twocolumn,
superscriptaddress,
amsmath,amssymb,
aps,
]{revtex4-2}

\usepackage{graphicx}% Include figure files
\usepackage{dcolumn}% Align table columns on decimal point
\usepackage{bm}% bold math
\usepackage{lipsum}
\usepackage[utf8]{inputenc}
\usepackage{enumitem}
\usepackage{soul} % nel preambolo
\usepackage{xcolor}

\begin{document}

\title{Enhanced sensitivity of sub-THz thermomechanical bolometers exploiting vibrational nonlinearity}

\author{L. Alborghetti}
\affiliation{Department of Physics, University of Pisa, Largo B. Pontecorvo 3, 56127 Pisa - Italy}
\author{B. Bertoni} 
\affiliation{Department of Physics, University of Pisa, Largo B. Pontecorvo 3, 56127 Pisa - Italy}
\affiliation{NEST, CNR Istituto Nanoscienze, piazza San Silvestro 12, 56127 Pisa - Italy}
\author{L. Vicarelli}
\affiliation{Department of Physics, University of Pisa, Largo B. Pontecorvo 3, 56127 Pisa - Italy}
\author{S. Zanotto}
\affiliation{NEST, CNR Istituto Nanoscienze, piazza San Silvestro 12, 56127 Pisa - Italy}
\author{S. Roddaro}
\affiliation{Department of Physics, University of Pisa, Largo B. Pontecorvo 3, 56127 Pisa - Italy}
\affiliation{NEST, CNR Istituto Nanoscienze, piazza San Silvestro 12, 56127 Pisa - Italy}
\author{A. Tredicucci}
\affiliation{Department of Physics, University of Pisa, Largo B. Pontecorvo 3, 56127 Pisa - Italy}
\affiliation{NEST, CNR Istituto Nanoscienze, piazza San Silvestro 12, 56127 Pisa - Italy}
\author{M. Cautero}
\affiliation{Department of Physics, Università degli Studi di Trieste, Piazzale Europa 1, 34127 Trieste, Italy}
\affiliation{Elettra - Sincrotrone Trieste S.C.p.A., Strada Statale 14, km 163.5, 34149 Trieste, Italy}
\author{L. Gregorat}
\affiliation{Department of Engineering and Architecture, Università degli Studi di Trieste, Via Alfonso Valerio 6/1, 34127 Trieste, Italy}
\affiliation{Elettra - Sincrotrone Trieste S.C.p.A., Strada Statale 14, km 163.5, 34149 Trieste, Italy}
\author{G. Cautero}
\affiliation{Elettra - Sincrotrone Trieste S.C.p.A., Strada Statale 14, km 163.5, 34149 Trieste, Italy}
\author{M. Cojocari}
\affiliation{Department of Physics and Mathematics, Center of Photonics Sciences, University of Eastern Finland, Yliopistokatu 7, FI-80101 Joensuu, Finland}
\author{G. Fedorov}
\affiliation{Department of Physics and Mathematics, Center of Photonics Sciences, University of Eastern Finland, Yliopistokatu 7, FI-80101 Joensuu, Finland}
\author{P. Kuzhir}
\affiliation{Department of Physics and Mathematics, Center of Photonics Sciences, University of Eastern Finland, Yliopistokatu 7, FI-80101 Joensuu, Finland}
\author{A. Pitanti}
\affiliation{Department of Physics, University of Pisa, Largo B. Pontecorvo 3, 56127 Pisa - Italy}
\affiliation{NEST, CNR Istituto Nanoscienze, piazza San Silvestro 12, 56127 Pisa - Italy}

\begin{abstract} 
A common approach to detecting weak signals or minute quantities involves leveraging the localized spectral features of resonant modes, whose sharper lines (i.e. high Q-factors) enhance transduction sensitivity. However, maximizing the Q-factor often introduces technical challenges in fabrication and design. In this work, we propose an alternative strategy to achieve sharper spectral features by using interference and nonlinearity, all while maintaining a constant dissipation rate. Using far-infrared thermomechanical detectors as a test case, we demonstrate that signal transduction along an engineered response curve slope effectively reduces the detector's noise equivalent power (NEP), achieving $\mathrm{\sim 30 \, pW/\sqrt{Hz}}$ NEP for electrical read-out, sub-THz detectors with an optimized absorbing layer. 
\end{abstract}

\maketitle
%%%%%%%%%%%%%%%%%%%%%%%%%%  body  %%%%%%%%%%%%%%%%%%%%%%%%%%
\section{Introduction}
Transducer-based sensing -- relying on the conversion of energy from one form (the measured quantity) to another (read-out signal) -- strongly benefits from spectrally sharp transfer functions, as large derivatives translate into large responsivities.
The usual route for optimizing the transfer functions relies on devising resonant elements with low energy loss rates; this manifests into narrow linewidths (large Q-factors) and leads to large sensitivities, which have been widely and successfully employed among the others in mass \cite{ikehara2007}, aerostatic pressure \cite{yang2016}, gas \cite{chen2014}, temperature \cite{sanchez2024}, polarization state \cite{zanotto2020} and refractive index \cite{yang2015} sensing.\\ Increasing the Q-factors of nano- and micro-metric sized detectors, with footprints suitable for integration in electronic systems, represents a significant technological challenge, both in terms of design and fabrication. Noteworthy, an intense research has recently led to ultra-high Q-factors (exceeding one billion) in micromechanical resonators oscillating at frequencies from hundreds of kHz to MHz, obtained through soft clamping and dissipation dilution design as well as careful fabrication \cite{hoj2021,tsaturyan2017,engelsen2024,cupertino2024}. Although impressive, reaching consistently these values in commercially compatible processes remains a significant challenge: device imperfections or intrinsic physical effects often create bottlenecks, ultimately limiting the maximum sensitivity of commercial devices, which routinely have Q-factors ranging around 10$^4$-10$^5$.\\ A different, interesting route to obtain sharp features in the transfer functions relies on manipulating the resonant lineshape, creating steeper edges while mantaining the same energy loss rate. To this end, Fano-induced asymmetry has been used for sensing enhancement, showing promising results for gas \cite{zaki2022, sherif2023, zaki2020}, refractive index \cite{zhu2021, tang2017} and temperature \cite{kong2017} sensing.\\ Improvements based on exploiting Fano lineshapes are still restricted by the limited manipulation of the Fano factor, which usually depends on the coupling between a broad and a narrow linewidth cavity system \cite{limonov2017}. Moreover, the maximum slope of a Fano response function is still connected to the resonator's Q-factor. This further limit can be surpassed by taking advantage of intrinsic device nonlinearities, which can produce massive linewidth deformations (foldover effect \cite{yang2024}) or lead to multistability\cite{pham2020}, with internal transitions between stable solutions leading to, in principle, ``infinitely sharp" spectral features.\\ These concepts have been successfully implemented in several systems including superconductors \cite{esmaeil2021} and semiconductors \cite{pfenning2022,blakesley2005} single photon detectors. The same ideas have also found focused applications in micromechanical systems, including mass sensors monitoring the shift of nonlinear resonant frequencies \cite{dai2009,dai2012}, ``threshold-based'' sensors exploiting bifurcation phenomena for mass \cite{kumar2012,yuksel2019} and gas \cite{al-ghamdi2018,shama2024} limit detection, as well as more structured platforms exploiting exceptional points \cite{zhang2024}.\\
\hfill\break
%***FIG 1*************************** 
\begin{figure*}[t]
	\centering
	\includegraphics[width=1\linewidth]{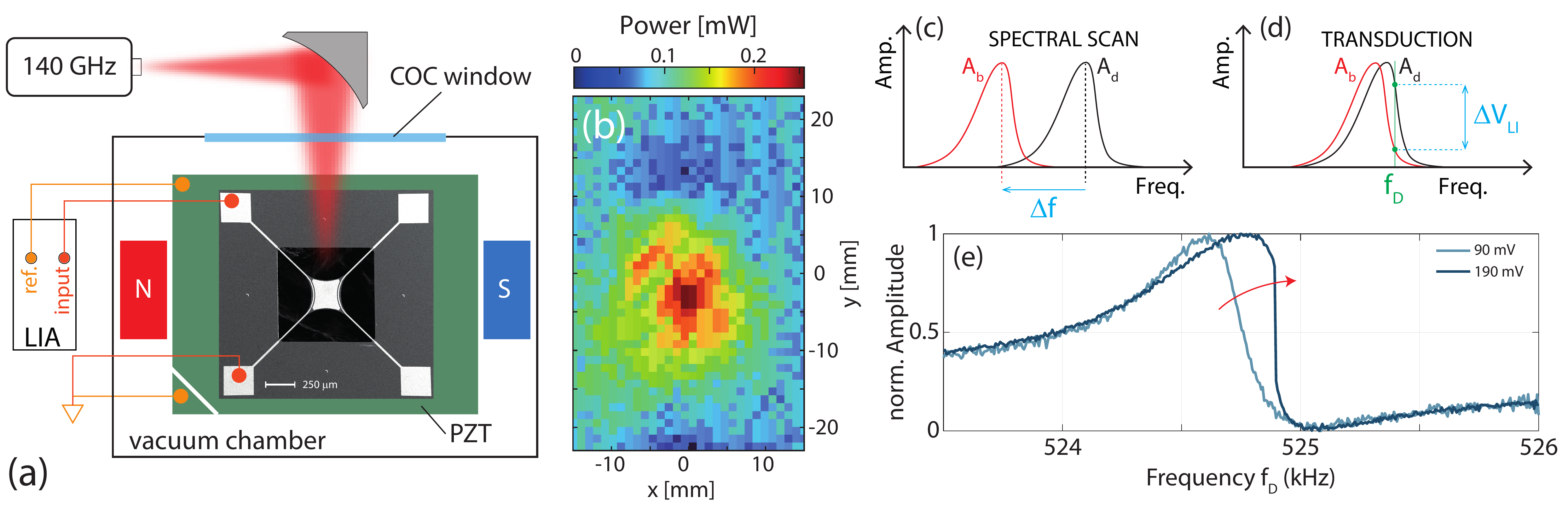}
	\caption{(a): Sketch of the experimental setup along with a SEM micrograph of one typical TMB. The device sits on a piezoelectric actuator stack (PZT) in a vacuum chamber with optical access via a cyclic olefin copolymer (COC) window. Static magnets generate a planar magnetic field which is used for the coherent magneto-motive read-out demodulated by a lock-in amplifier (LIA) at the forcing mechanical drive. (b): Beam profiling of the 140 GHz source. Comparison of the Spectral Scan (c) and Transduction (d) operating modes. (e): Distortion of the resonance lineshape of the trampoline resonator by increasing the driving amplitude and entering strong nonlinear motional regimes.}
	\label{fig:1}
\end{figure*}
In this paper, we show how nonlinear transduction sensing can be employed to enhance the characteristics of far-infrared light detectors based on thermo-mechanical bolometers (TMBs). TMBs have recently emerged as a powerful system which offers broadband detection at room-temperature, with single-pixel operation at video-rate and faster \cite{blaikie2019,zhang2019,vicarelli2022,piller2022,zhang2024b} and the possibility of scaling up the system to focal plane arrays for multiplexed imaging applications \cite{duraffourg2018,gregorat2024}. The best TMB devices have a noise-equivalent power (NEP) in a range from a few $\mathrm{pW/\sqrt{Hz}}$ level, to some  $\mathrm{nW/\sqrt{Hz}}$, in some cases outperforming commercially available technologies \cite{li2023}.\\ Our approach to TMBs makes use of silicon nitride (Si$_3$N$_4$) trampoline resonators, which have previously shown a NEP of about  $\mathrm{100 \, \, pW/\sqrt{Hz}}$  at 20 Hz operating speed in a 1$\times$1 mm sided membrane, illuminated with a 140 GHz source and optically read-out via self-mixing interferometry \cite{vicarelli2022}.\\
Compared to our previous results, the devices here investigated have a reduced pixel size and employ specific layers for improved absorptance. In addition, addressing the TMB with metallic wires in a magnetic field, we switched to all-electrical probing via inductive read-out \cite{schmid2016}, which is better suited for integration and parallelization in large-scale arrays, with the trade-off of having a generically higher noise floor.\\ 
By applying nonlinear transduction schemes, based on the material Duffing nonlinearity \cite{Schuster2008}, we achieve a NEP of about $\mathrm{30 \, \, pW/\sqrt{Hz}}$, evaluated under an illuminating 140 GHz source and at room temperature. Combined with an operating speed of 20 Hz, our device characteristics and measurement technique challenge the state of the art for room-temperature bolometric detectors in the sub-THz range.\\
It is generally accepted that the high detector sensitivity comes at a price of reduced dynamic range. Extremely sensitive detectors are only used for extremely weak signals and cannot operate in a wide range of intensities due to saturation or even damage. In our case the degree of nonlinearity controlled through the amplitude of exciting voltage can be used to tune the detector from the regime of reasonable sensitivity combined with high dynamic range (low excitation voltage, linear regime) to the state of a high sensitivity required to detect very weak radiation.

%*********************************** 

\section{Methods}

\subsection{Device details}

The devices are based on Si$_3$N$_4$ trampoline membrane resonators, which in the last decade have been successfully employed for classical sensing \cite{kanellopulos2024} and quantum applications \cite{norte2016,reinhardt2016}. The basic device geometry consists of single membranes made of  a 300 nm thick stoichiometric silicon nitride, with a 100$\times$100 $\mathrm{\mu}$m central plate hanging on a 300$\times$300 $\mathrm{\mu}$m frame through four 12 $\mathrm{\mu m}$ wide tethers. All membranes have Cr/Au metallic contacts running through the tethers. These grant a practical all-electrical read-out and actuation \cite{venstra2009,schmid2016,chien2020}, enabled by a $250$ mT magnetic field induced by static Nd magnets and leveraging inductive reading or Lorentz force, respectively. A scanning electron micrograph of a typical device is reported in \textbf{Figure~\ref{fig:1}} (a) along with a sketch outlining the read-out circuit.\\
In order to enhance the device absorbance without degrading its mechanical properties, we introduced ``ultra-light" bidimensional layers in the center of the membrane. We have explored the implementation of both metallic and carbon-based absorbers  \cite{lamura2023}. The choice for  metal was a Cr/Au bilayer with an overall thickness of about 8 nanometers. At such a small scale the metal layer will be a non-uniform film composed by adjacent grains, increasing its sheet resistance compared to its bulk metal value. This favored approaching the limit value of 188 $\Omega$ sheet resistance which would nominally give perfect impedance matching and the theoretical 50\% absorption limit for an isolated layer \cite{Haddadi2022}. The other avenue considered makes use of amorphous carbon materials which can be directly grown on silicon nitride via chemical vapour deposition \cite{kaplas2017, kaplas2012}, allowing fine adjustments of the layer conductivity through the precise control of the film thickness. The material of our choice, pyrolitic carbon (PyC), has shown an impressive absorption of $\sim$ 43\% in the sub-THz range \cite{jorudas2024}. 
We realized and investigated two different device implementations, with the same nominal geometrical parameters for the trampoline resonator and different absorbing layers, namely a 2/6 nm Cr/Au thick layer (\textbf{Au device}) and a 18 nm pyrolitic carbon  thick layer (\textbf{PyC device}), respectively.\\

\subsection{Detector read-out}

The devices were mounted on a ceramic piezoeletric stack actuator layer which was used to coherently excite the membrane motion (see \textbf{Figure~\ref{fig:1}} (a)). The AC drive voltage applied on the actuator was generated by the reference channel of a lock-in amplifier (LIA), whose input channel was connected to the TMB contact to sense the magnetomotive voltage $\mathrm{\Delta V}$, induced by the modulation of the concatenated magnetic flux due to the vibrational displacement. TMB, actuator and static magnets were placed on a custom-made printed circuit board; this was inserted in a vacuum chamber  to reduce atmospheric viscous damping. Optical access was enabled through a cyclic-olefin-copolymer (COC) window. The vacuum chamber was additionally mounted on a planar moving stage ($xy$ plane), used to scan the TMB position for extended imaging. The illumination was done via a continuous wave, 30 mW, 140 GHz source focused on the device with two parabolic mirrors (not shown in the sketch of \textbf{Figure~\ref{fig:1}} (a)).\\
\hfill\break
The concept behind the detection mechanism lies in the shift of the resonant frequency of specific mechanical modes due to thermally induced device deformations, namely thermal expansion and tensile stress reduction. The temperature change is accordingly caused by the absorption of the electromagnetic radiation which one wants to detect. Sweeping the driving frequency, it is possible to directly acquire the whole mechanical spectrum and evaluate the frequency shift $\Delta f$ of specific features of the ``bright" spectrum ($A_b$) from the ``dark" ($A_d$) one (Spectral Scan, see \textbf{Figure~\ref{fig:1}} (c)). In our operating conditions, the frequency shift scales linearly with the illuminating intensity, allowing the use of this operating mode with large dynamic ranged signals as well as for the direct acquisition of images, as reported, for example, in \textbf{Figure~\ref{fig:1}} (b), where the source focused beam profile has been imaged via spatial scan. The linear response allows a direct conversion of the image from frequency shift to impinging power; here this was done by imposing the proper beam normalization to its total power, which has been independently measured through a calibrated Golay cell detector, taking also into account the absorption of the COC window, which at this frequency stands around $50\%$.\\
Note that the spectral scan is generally a slow detection method, often limited by the acquisition time of lock-in amplifiers due to the long frequency sweeps, especially when multiple devices are simultaneously investigated.\\ 
Faster detection protocols with a reduced dynamic range rely instead on operating with a single frequency driving/demodulating tone $f_D$ and exploiting the transduction effect in an open-loop configuration, as illustrated in the transduction scheme of \textbf{Figure~\ref{fig:1}} (d). 
In our device, the transduction detection speed can reach video-rate \cite{vicarelli2022}, limited by the thermal relaxation time through the tethers. This process, with Q-factors for our devices ranging between $10^3$ and $10^5$, fully dominates the dynamics of the entire TMB upon illumination. In the transduction scheme, the frequency shift is directly converted into a read-out voltage $\Delta V_{LI}$ which depends on the difference between dark and bright spectrum amplitudes at $f_D$, $A_d(f_D)$ and $A_b(f_D)$, respectively. 
For the detection of weak signals, the bright spectrum can be recast as a frequency shift of the dark spectrum by a vanishing $\delta f$, obtaining:
\begin{equation}
\Delta V_{LI} \sim A_d(f_D)-A_d(f_D+\delta f)\sim \left.\frac{dA_d}{df}\right|_{f=f_D} \delta f
\end{equation} 
and giving a read-out voltage directly proportional to the first derivative of the spectral amplitude at $f_D$. Changing $f_D$ allows one to explore different regions of the signal derivative, which can be very large in asymmetric and nonlinear resonances. In particular, our resonance shows both Fano interference and nonlinear hardening due to the Duffing effect, which is known to produce an increase in the steepness of the spectral features. This is illustrated in \textbf{Figure~\ref{fig:1}} (e), which displays typical resonance lineshapes of a \textbf{PyC device} at different driving strengths: one can see that the already asymmetric Fano resonance becomes steeper around 525 kHz due to the Duffing effect for the larger driving voltage amplitude.

\section{Spectral transduction}

%***FIG 2***************************
\begin{figure}[t]
  \centering
  \includegraphics[width=0.45\textwidth]{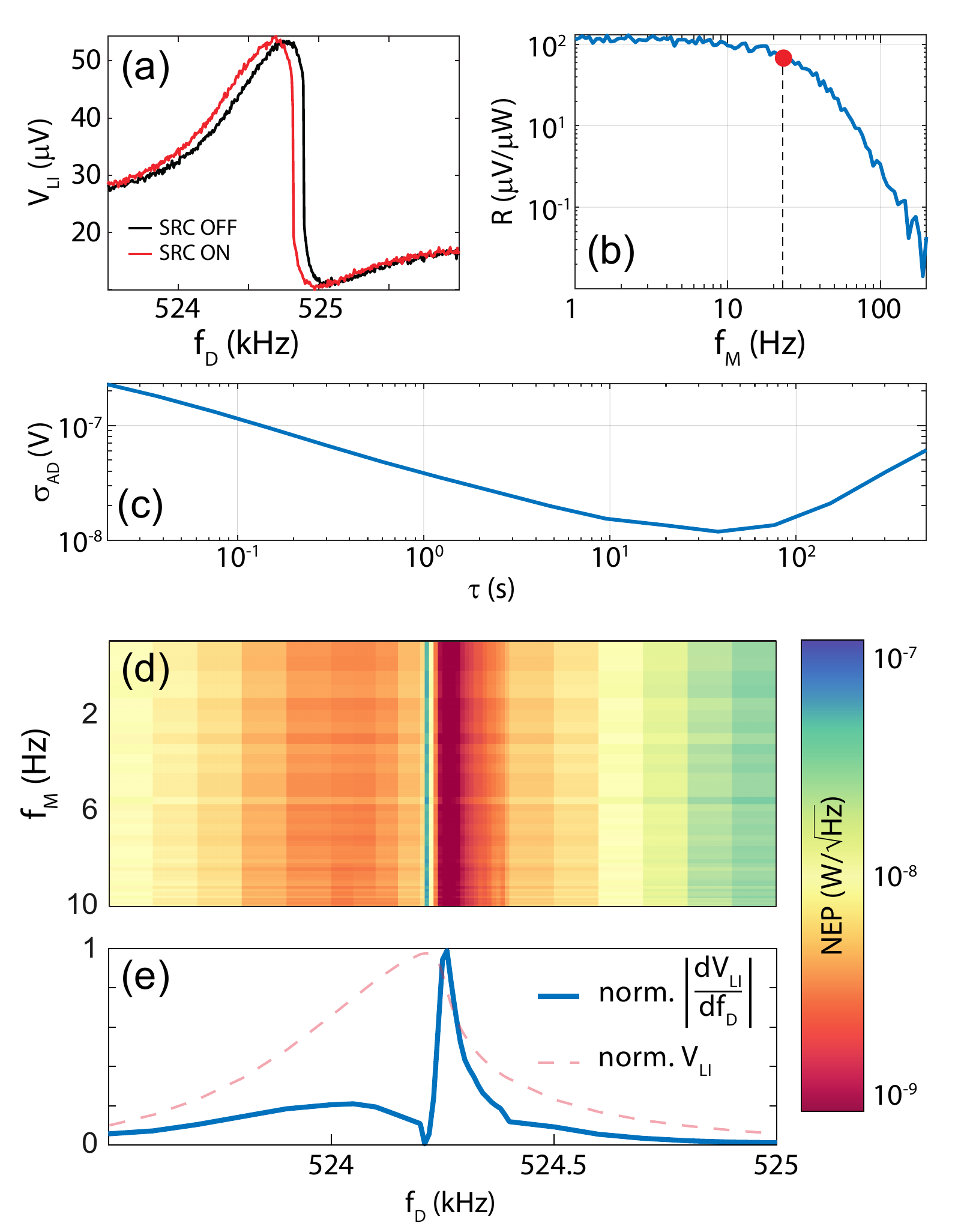}
  \caption{(a): typical ON/OFF spectra for the \textbf{PyC device} with a 190 mV piezo driving voltage. (b): dynamic responsivity for the \textbf{PyC device}. The red dot indicates the cut-off frequency. (c): Allan deviation for the \textbf{PyC device} with a demodulation bandwidth of $200 \, \mathrm{Hz}$. (d): spectrogram of PyC device NEP. The minimum NEP can be found around 524.26 kHz. (e): normalized derivative and corresponding OFF spectrum (dashed) for the \textbf{PyC device} with a 90 mV piezo driving voltage. }
  \label{fig:2}
\end{figure}
%***********************************

We evaluated the effect of asymmetric lineshapes on the figure of merit of transduction experiments with our TMB devices. Typical amplitude spectra of the \textbf{PyC device}, demodulated with a sweeping mechanical driving voltage of 190 mV and with on/off source, respectively, are shown in \textbf{Figure~\ref{fig:2}}.\\
Comparing dark and bright spectrum we can extract the device static responsivity, using the calibrated intensity map of \textbf{Figure~\ref{fig:1}} (b) for estimating the radiation power illuminating the detector. Although one could consider the power deposited on the whole membrane area (300 $\mathrm{\mu m}\times$300 $\mathrm{\mu m}$) or the diffraction-limited area $\lambda^2/4\approx 1.15 \, \mathrm{mm^2}$, for the sake of comparison we decided to adopt the same methodology commonly used in the literature \cite{Martini2025} and considered the power illuminating the absorbing layer region ($60 \, \mathrm{\mu m}$ x $85 \, \mathrm{\mu m}$). This yields an illuminating power of $P_{i}\sim 0.26 \, \mathrm{\mu W}$ and $\sim 0.48 \, \mathrm{\mu W}$ for the \textbf{PyC} and \textbf{Au} device, respectively. The different illumination power is due to the slightly different $xy$ positions where we placed the TMBs.   
 
Operating the device in transduction mode and considering a weak signal approximation, a static responsivity can be defined for each driving frequency $f_D$ as:
\begin{equation}
\label{eq:resp}
R_s(f_D)\sim \left.\frac{d V_{LI}}{d f}\right|_{f=f_D} \left(\frac{\Delta f}{P_i}\right)
\end{equation}
where the static spectral shift per unit power $\Delta f/P_i$ can be estimated from the results of \textbf{Fig \ref{fig:2}} (a). The dynamic responsivity $R$ was then evaluated by combining the previous results with the Bode frequency response, obtained using square wave modulation of the source in a frequency range $f_M$ from 1 to 120 Hz, with an external electrical signal provided by the LIA. A typical dynamic responsivity measured at a driving frequency of 524.8 kHz is reported in \textbf{Figure~\ref{fig:2}} (b). As expected, the thermal response of the membrane acts as a low-pass filter, with cut-off frequency given by the inverse of thermal relaxation time $\tau_{th}$. The cut-off frequency extracted from the Bode is about 20 Hz, as indicated by a red dot in \textbf{Figure~\ref{fig:2}} (b), and corresponds to a $\tau_{th}\sim50$ ms, compatible with other reports in the literature \cite{gregorat2024, Martini2025}.\\ 
Next, the device noise was measured in dark conditions, by calculating the Allan deviation $\sigma_{AD}$, which provides a time-domain characterization of the resonator frequency stability. The Allan deviation was computed from signal recordings over one minute at various driving frequencies, using the UHF Lock-in amplifier (Zurich Instruments), with demodulation bandwidth of $10 \, \mathrm{Hz}$, which lies below the thermal cut-off frequency of the device. A representative measurement from a longer acquisition is shown in \textbf{Figure~\ref{fig:2}}(c), using a demodulation bandwidth of $200 \, \mathrm{Hz}$ to show the behavior at lower $\tau$. The typical behavior is observed: at short averaging times $\tau$, the Allan deviation follows a $\propto \tau^{-1/2}$ trend, indicative of white frequency noise dominance, while at longer $\tau$, the impact of frequency drift becomes evident.

Allan deviation and dynamic responsivity can be combined to yield the Noise-Equivalent Power; considering the time-dependence of both  the Bode plot of Fig. \ref{fig:2} (b) and the noise contribution of Fig. \ref{fig:2} (c), as expected, the NEP is a function of the frequency of modulation of the incoming signal ($f_M$). Moreover, given the strong asymmetry of our lineshapes, in our device the NEP is also function of the mechanical driving frequency $f_D$:
\begin{equation}\label{eq:NEP}
NEP(f_M,f_D) = \frac{\sigma_{AD} \sqrt{2 \tau}}{R}
\end{equation}
The full NEP spectrogram for the \textbf{PyC device} excited with a 90 mV driving tone can then be expressed in the $f_M-f_D$ plane, as reported in \textbf{Figure~\ref{fig:2}} (d). It is significant to observe that there is a drastic reduction of the NEP at a driving frequency of $f_D$=524.26 kHz, reaching values below $1\;\mathrm{nW/\sqrt{Hz}}$, significantly reduced with respect to the rest of the spectrum. Unsurprisingly, this driving frequency corresponds to the spectral region with the highest slope, as can be seen from \textbf{Figure~\ref{fig:2}} (e), where we reported the normalized modulus of the derivative calculated from the resonance lineshape (also reported as a dashed line).\\
Qualitatively similar effects are present at different driving voltages: higher drives lead to even sharper lineshapes, with a corresponding decrease in the NEP in narrow spectral regions. Conversely, lower drives produce an increase in the NEP, albeit showing the best performances in a broader spectral range. This concept is directly linked to the device dynamic range of operation: large spectral regions with significant slopes are most beneficial for using the detector for imaging or as a power meter; at the other limit, extremely large slope in very narrow frequency ranges, we expect the device to operate as a threshold-switching detector, better suited for the recognition of single events (i.e. laser pulses), as will be discussed later.

%***FIG 3***************************
\begin{figure}[t]
  \centering
  \includegraphics[width=0.45\textwidth]{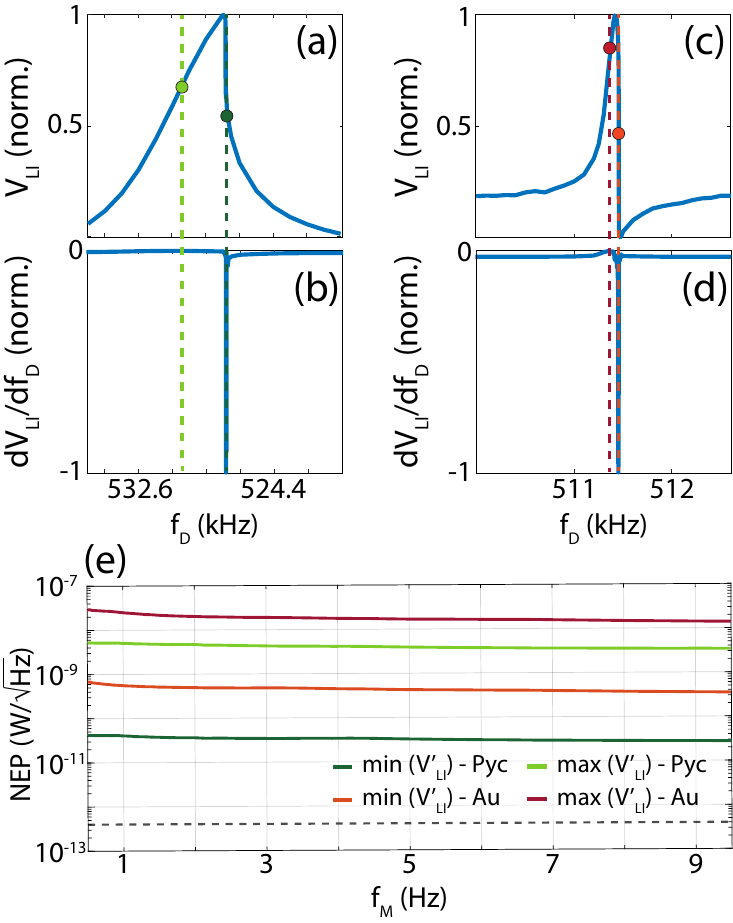}
  \caption{Normalized vibrational spectrum (a) and its first derivative (b) for the PyC device. Normalized vibrational spectrum (c) and its first derivative (d) for the Au device. The dashed lines indicate in both cases the position of the maximum and minimum derivative, respectively. (e): NEP evaluated for $f_D$ corresponding to the maximum (i.e., most positive) and minimum (i.e., most negative) of the derivative for both devices. The dashed line indicates the theoretical thermal limit of the NEP, calculated for the \textbf{PyC device}.}
  \label{fig:3}
\end{figure}
%***********************************
\hfill\break  
As we have illustrated, a critical parameter in our operating scheme is the derivative of the mechanical resonance: changing the transduction frequency can dramatically change the device response. For example, \textbf{Figure~\ref{fig:3}} (a) shows the normalized mechanical spectrum of \textbf{PyC device} under a driving voltage of 112 mV.

The asymmetry becomes even more manifest by looking at the derivative: the largest negative slope is about 5.6 times the largest positive one, see the green dots and lines in \textbf{Figure~\ref{fig:3}} (b). The \textbf{Au device} shows a qualitatively similar asymmetry, as shown in \textbf{Figure~\ref{fig:3}} (c-d). The two devices are nominally identical apart from the different absorption layer. Differences in their resonant spectra can be ascribed to the different experimental assembly, starting from the local properties of the piezoelectric actuator where they are glued upon. The piezoelectric stack represents a cross-talk channel strongly impacting the Fano factor, see a more detailed modeling in \cite{gregorat2024}. For a fair device comparison, we then evaluated the driving voltage leading to the bifurcation point (discussed in more detail later): this acts as a common reference, since we expect both devices to have the same vibrational amplitude at bifurcation. Scaling from this value, we applied a driving voltage of 118 mV to the \textbf{Au device} (\textbf{Figure~\ref{fig:3}} (c-d)), so that it can be compared to the 112 mV drive of the PyC one (\textbf{Figure~\ref{fig:3}} (a-b)). The \textbf{Au device} also exhibits a strong asymmetry (roughly a 37.5 times increase of the negative to the positive slope) and generally a comparable but smaller derivative with respect to the PyC device.\\

The NEP evaluated by setting $f_D$ at the maximum and minimum slopes of both devices is reported in \textbf{Figure~\ref{fig:3}} (e) as a function of the modulation frequency of the 140 GHz source. All NEP curves remain relatively constant as a function of $f_M$, consistent with the results of \textbf{Figure~\ref{fig:2}} (d), and show an overall low value which strongly depends of the chosen transduction frequency. As expected from Eqs. (\ref{eq:NEP}) and (\ref{eq:resp}), the NEP is inversely proportional to the derivative, resulting in net reductions of about a factor of 117 and 37.5, respectively, in good agreement with what one would expect from the results of Fig. \ref{fig:3} (b) and (d). Note that the NEP reductions are obtained within the same mechanical device, and obtained just by changing $f_D$. 
Moreover, the \textbf{PyC device} shows about an order of magnitude improvement with respect to the \textbf{Au device} which originates from the different absorbance of the two materials at 140 GHz. Interestingly, comparing the best NEP values in the two devices, appropriately re-scaled to account for the different slopes, gives an absorption-induced enhancement of $\sim$7.94, in good agreement with the ratio of static responsivities, evaluated as $R_s^{PyC}\sim$497 MHz/W and $R_s^{Au}\sim$72.6 MHz/W, giving $R_s^{PyC}/R_s^{Au}=6.85$. From this observation, we can estimate the absorbance of the metallic layer, starting from an experimentally measured PyC film absorbance of $\sim$40\% in the far-infrared range \cite{jorudas2024}. The obtained Cr/Au absorbance of $\sim$10\% is compatible with multilayer simulations using the experimentally estimated refractive indices of Si$_3$N$_4$ and metals \cite{ordal1983}, which returns a numerical value of $4\pm0.5\%$ for a continuous film, whereas we can assume a further enhancement due to the layer inhomogeneity.\\
	
The best NEP we obtained is about 30 $\mathrm{pW/\sqrt{Hz}}$ for the \textbf{PyC device}. As reported in \textbf{Figure~\ref{fig:3}}(e), we still have room towards the fundamental NEP limit for a detector operating at a temperature $\mathrm{T}$, set by thermal fluctuations in the system and defined as \cite{rogalski2010infrared}:
\begin{equation}
	NEP^* = \sqrt{16k_BT^5\sigma A}
\end{equation}
where $k_B$ is the Boltzmann constant, $\sigma$ is the Stefan–Boltzmann constant and $\mathrm{A}$ is the absorber area. Even if some thermal detectors shows exceptional performances and metrics close to the thermal limit in the mid-infrared range, they rely on all-optical probes, which are known to inject less noise to the system with respect to the all-electrical ones employed here \cite{Martini2025}. Conversely, all-electrical read-out is better suitable for portability and integration and our NEP compares well with some of the best commercial devices in the sub-THz range, which have NEPs around 10 $\mathrm{pW/\sqrt{Hz}}$ \cite{VDI}. 
\hfill\break
The main drawback of our technique lies in the reduced bandwidth where we find very large derivatives. This directly translates into a reduction of the detector dynamic range (i.e. the ratio between the maximum and minimum detectable power). As a quantitative example, with the driving condition of \textbf{Figure~\ref{fig:3}}, the \textbf{PyC device} negative derivative peak has a linewidth of about 50 Hz, limiting its use to weak signals with power less than roughly 100 nW, as estimated by considering the spectral static device responsivity.
\hfill\break

\begin{figure}[t]
  \centering
  \includegraphics[width=0.45\textwidth]{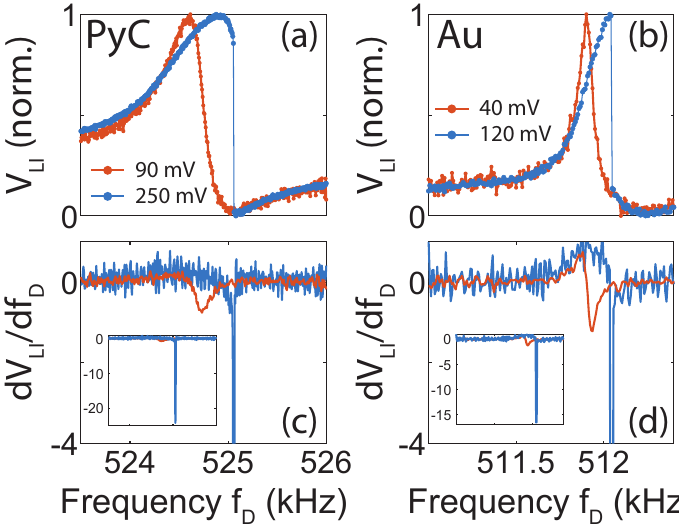}
  \caption{Normalized vibrational spectrum (a) of the PyC device under a weak (90 mV) and a strong (250 mV) driving voltage amplitude, respectively. The large drive amplitude leads to a multistable regime where the oscillator jumps from one stable solution to the other. The multistability leads to delta-like responsivity, as seen in the zoomed-in calculated first derivative of (c) (inset: full scale derivative comparison). (b,d): Similar measurements for the Au device.}
  \label{fig:4}
\end{figure}

A further increase of the driving voltage leads the system to highly nonlinear regimes which, in the case of Duffing nonlinearities, can eventually reach multistable or chaotic motion \cite{Schuster2008}. As an example, \textbf{Figure~\ref{fig:4}} (a-b) reports the comparison of an asymmetric, single-solution resonance (red curve) and a regime past the bifurcation point (blue curve) for both devices under investigation. The multistable regime is characterized by abrupt jumps resulting from switching between two stable solutions; the onset of multistability essentially depends on the vibrational amplitude, since the Duffing nonlinearity enters the equation of motion with a cubic displacement term \cite{gregorat2024}. The bifurcation point from a single to multiple dynamical solutions has been taken as a common reference to compare both devices.\\ 
One can think of the switching as a ``line with infinite slope'', which looks appealing for transduction detection. Unfortunately, in this case the derivative tends to a $\delta$-function, as can be seen from the normalized numerical derivates reported in \textbf{Figure~\ref{fig:4}} (c-d), respectively. Evaluating the figures (and the insets for the full range of the $y$ axis), one can expect an extreme enhancement of device performance in terms of responsivity, which comes along with a vanishing dynamic range. 

%***FIG 4***************************
%***********************************

Since diverging derivatives can be found at a single frequency point, this particular regime of operation does not allow utilizing the device as an intensity detector; conversely it looks promising for a threshold-based sensor, where the discontinuity in the read-out can be triggered by specific events over a certain limit (i.e. single photons, light pulse detection, temperature change, mass loading). In a scheme similar to the superconducting optical detectors \cite{esmaeil2021} or bifurcation-based sensors \cite{kumar2012,yuksel2019,al-ghamdi2018,shama2024,zhang2024} in other nonlinear systems, this operating regime adds to the potential and possibilities lying in thermomechanical bolometer devices.\\

\section{Conclusion}

While increasing the quality factor in resonant detectors generally enhances device performance, consistently achieving high values can be challenging, especially in the context of mass production for real-world applications. In this work, we presented an alternative approach to locally achieve high responsivity and low NEP in transduction detection experiments by exploiting device nonlinearities.
Through careful characterization of thermomechanical bolometers, we demonstrated that the device NEP under 140 GHz illumination is strongly dependent on the driving strength and transduction frequencies. In the optimal operating range, we achieved a NEP of 30 $\mathrm{pW/\sqrt{Hz}}$ for a TMB employing pyrolitic carbon as an absorbing layer, which exhibits a 40\% absorbance in the sub-THz range under investigation \cite{jorudas2024}. 
Additionally, by further increasing the driving strength and entering a highly nonlinear regime, we propose leveraging the same platform for threshold signal detection. In this regime, the system can transition between stable solutions in response to external perturbations of sufficient magnitude, with a behaviour similar to superconducting or bifurcation-based detectors, opening up further possibilities for our technology.

\section{Acknowledgment}
The authors gratefully acknowledge funding from ATTRACT, a European Union’s Horizon 2020 research and innovation project under grant agreement No. 101004462 (H-cube project), from European Union - Next Generation EU under the Italian National Recovery and Resilience Plan (NRRP), Mission 4, Component 2, Investment 1.3, CUP D43C22003080001, partnership on “Telecommunications of the Future” (PE00000001 - program “RESTART”) and from MUR through the project PRIN 2022 TRUST. This work is supported by the Academy of Finland via Flagship Programme Photonics Research and Innovation (PREIN), decision 368653, Horizon 2020 RISE CHARTIST project 101007896, Horizon 2020 RISE TERASSE project 823878.

%%%%%%%%%%%%%%%%%%%%%%% References %%%%%%%%%%%%%%%%%%%%%%%%%

\bibliography{nonlinear_detect}

%apsrev4-2.bst 2019-01-14 (MD) hand-edited version of apsrev4-1.bst
%Control: key (0)
%Control: author (8) initials jnrlst
%Control: editor formatted (1) identically to author
%Control: production of article title (0) allowed
%Control: page (0) single
%Control: year (1) truncated
%Control: production of eprint (0) enabled
\begin{thebibliography}{53}%
\makeatletter
\providecommand \@ifxundefined [1]{%
 \@ifx{#1\undefined}
}%
\providecommand \@ifnum [1]{%
 \ifnum #1\expandafter \@firstoftwo
 \else \expandafter \@secondoftwo
 \fi
}%
\providecommand \@ifx [1]{%
 \ifx #1\expandafter \@firstoftwo
 \else \expandafter \@secondoftwo
 \fi
}%
\providecommand \natexlab [1]{#1}%
\providecommand \enquote  [1]{``#1''}%
\providecommand \bibnamefont  [1]{#1}%
\providecommand \bibfnamefont [1]{#1}%
\providecommand \citenamefont [1]{#1}%
\providecommand \href@noop [0]{\@secondoftwo}%
\providecommand \href [0]{\begingroup \@sanitize@url \@href}%
\providecommand \@href[1]{\@@startlink{#1}\@@href}%
\providecommand \@@href[1]{\endgroup#1\@@endlink}%
\providecommand \@sanitize@url [0]{\catcode `\\12\catcode `\$12\catcode
  `\&12\catcode `\#12\catcode `\^12\catcode `\_12\catcode `\%12\relax}%
\providecommand \@@startlink[1]{}%
\providecommand \@@endlink[0]{}%
\providecommand \url  [0]{\begingroup\@sanitize@url \@url }%
\providecommand \@url [1]{\endgroup\@href {#1}{\urlprefix }}%
\providecommand \urlprefix  [0]{URL }%
\providecommand \Eprint [0]{\href }%
\providecommand \doibase [0]{https://doi.org/}%
\providecommand \selectlanguage [0]{\@gobble}%
\providecommand \bibinfo  [0]{\@secondoftwo}%
\providecommand \bibfield  [0]{\@secondoftwo}%
\providecommand \translation [1]{[#1]}%
\providecommand \BibitemOpen [0]{}%
\providecommand \bibitemStop [0]{}%
\providecommand \bibitemNoStop [0]{.\EOS\space}%
\providecommand \EOS [0]{\spacefactor3000\relax}%
\providecommand \BibitemShut  [1]{\csname bibitem#1\endcsname}%
\let\auto@bib@innerbib\@empty
%</preamble>
\bibitem [{\citenamefont {Ikehara}\ \emph {et~al.}(2007)\citenamefont
  {Ikehara}, \citenamefont {Lu}, \citenamefont {Konno}, \citenamefont {Maeda},\
  and\ \citenamefont {Mihara}}]{ikehara2007}%
  \BibitemOpen
  \bibfield  {author} {\bibinfo {author} {\bibfnamefont {T.}~\bibnamefont
  {Ikehara}}, \bibinfo {author} {\bibfnamefont {J.}~\bibnamefont {Lu}},
  \bibinfo {author} {\bibfnamefont {M.}~\bibnamefont {Konno}}, \bibinfo
  {author} {\bibfnamefont {R.}~\bibnamefont {Maeda}},\ and\ \bibinfo {author}
  {\bibfnamefont {T.}~\bibnamefont {Mihara}},\ }\bibfield  {title} {\bibinfo
  {title} {A high quality-factor silicon cantilever for a low detection-limit
  resonant mass sensor operated in air},\ }\href
  {https://doi.org/10.1088/0960-1317/17/12/015} {\bibfield  {journal} {\bibinfo
   {journal} {J. Micromech. Microeng.}\ }\textbf {\bibinfo {volume} {17}},\
  \bibinfo {pages} {2491} (\bibinfo {year} {2007})}\BibitemShut {NoStop}%
\bibitem [{\citenamefont {Yang}\ \emph {et~al.}(2016)\citenamefont {Yang},
  \citenamefont {Saurabh}, \citenamefont {Ward},\ and\ \citenamefont
  {Nic~Chormaic}}]{yang2016}%
  \BibitemOpen
  \bibfield  {author} {\bibinfo {author} {\bibfnamefont {Y.}~\bibnamefont
  {Yang}}, \bibinfo {author} {\bibfnamefont {S.}~\bibnamefont {Saurabh}},
  \bibinfo {author} {\bibfnamefont {J.~M.}\ \bibnamefont {Ward}},\ and\
  \bibinfo {author} {\bibfnamefont {S.}~\bibnamefont {Nic~Chormaic}},\
  }\bibfield  {title} {\bibinfo {title} {High-q, ultrathin-walled microbubble
  resonator for aerostatic pressure sensing},\ }\href@noop {} {\bibfield
  {journal} {\bibinfo  {journal} {Optics express}\ }\textbf {\bibinfo {volume}
  {24}},\ \bibinfo {pages} {294} (\bibinfo {year} {2016})}\BibitemShut
  {NoStop}%
\bibitem [{\citenamefont {Chen}\ \emph {et~al.}(2014)\citenamefont {Chen},
  \citenamefont {Han}, \citenamefont {Liu},\ and\ \citenamefont
  {Hong}}]{chen2014}%
  \BibitemOpen
  \bibfield  {author} {\bibinfo {author} {\bibfnamefont {T.}~\bibnamefont
  {Chen}}, \bibinfo {author} {\bibfnamefont {Z.}~\bibnamefont {Han}}, \bibinfo
  {author} {\bibfnamefont {J.}~\bibnamefont {Liu}},\ and\ \bibinfo {author}
  {\bibfnamefont {Z.}~\bibnamefont {Hong}},\ }\bibfield  {title} {\bibinfo
  {title} {Terahertz gas sensing based on a simple one-dimensional photonic
  crystal cavity with high-quality factors},\ }\href
  {https://doi.org/10.1364/AO.53.003454} {\bibfield  {journal} {\bibinfo
  {journal} {Appl. Opt., AO}\ }\textbf {\bibinfo {volume} {53}},\ \bibinfo
  {pages} {3454} (\bibinfo {year} {2014})}\BibitemShut {NoStop}%
\bibitem [{\citenamefont {S{\'a}nchez-Pastor}\ \emph
  {et~al.}(2024)\citenamefont {S{\'a}nchez-Pastor}, \citenamefont
  {Kad{\u{e}}ra}, \citenamefont {Sakaki}, \citenamefont {Jakoby}, \citenamefont
  {Lacik}, \citenamefont {Benson},\ and\ \citenamefont
  {Jim{\'e}nez-S{\'a}ez}}]{sanchez2024}%
  \BibitemOpen
  \bibfield  {author} {\bibinfo {author} {\bibfnamefont {J.}~\bibnamefont
  {S{\'a}nchez-Pastor}}, \bibinfo {author} {\bibfnamefont {P.}~\bibnamefont
  {Kad{\u{e}}ra}}, \bibinfo {author} {\bibfnamefont {M.}~\bibnamefont
  {Sakaki}}, \bibinfo {author} {\bibfnamefont {R.}~\bibnamefont {Jakoby}},
  \bibinfo {author} {\bibfnamefont {J.}~\bibnamefont {Lacik}}, \bibinfo
  {author} {\bibfnamefont {N.}~\bibnamefont {Benson}},\ and\ \bibinfo {author}
  {\bibfnamefont {A.}~\bibnamefont {Jim{\'e}nez-S{\'a}ez}},\ }\bibfield
  {title} {\bibinfo {title} {A wireless w-band 3d-printed temperature sensor
  based on a three-dimensional photonic crystal operating beyond 1000° c},\
  }\href@noop {} {\bibfield  {journal} {\bibinfo  {journal} {Communications
  Engineering}\ }\textbf {\bibinfo {volume} {3}},\ \bibinfo {pages} {137}
  (\bibinfo {year} {2024})}\BibitemShut {NoStop}%
\bibitem [{\citenamefont {Zanotto}\ \emph {et~al.}(2020)\citenamefont
  {Zanotto}, \citenamefont {Tredicucci}, \citenamefont {Navarro-Urrios},
  \citenamefont {Cecchini}, \citenamefont {Biasiol}, \citenamefont
  {Mencarelli}, \citenamefont {Pierantoni},\ and\ \citenamefont
  {Pitanti}}]{zanotto2020}%
  \BibitemOpen
  \bibfield  {author} {\bibinfo {author} {\bibfnamefont {S.}~\bibnamefont
  {Zanotto}}, \bibinfo {author} {\bibfnamefont {A.}~\bibnamefont {Tredicucci}},
  \bibinfo {author} {\bibfnamefont {D.}~\bibnamefont {Navarro-Urrios}},
  \bibinfo {author} {\bibfnamefont {M.}~\bibnamefont {Cecchini}}, \bibinfo
  {author} {\bibfnamefont {G.}~\bibnamefont {Biasiol}}, \bibinfo {author}
  {\bibfnamefont {D.}~\bibnamefont {Mencarelli}}, \bibinfo {author}
  {\bibfnamefont {L.}~\bibnamefont {Pierantoni}},\ and\ \bibinfo {author}
  {\bibfnamefont {A.}~\bibnamefont {Pitanti}},\ }\bibfield  {title} {\bibinfo
  {title} {Optomechanics of chiral dielectric metasurfaces},\ }\href@noop {}
  {\bibfield  {journal} {\bibinfo  {journal} {Advanced Optical Materials}\
  }\textbf {\bibinfo {volume} {8}},\ \bibinfo {pages} {1901507} (\bibinfo
  {year} {2020})}\BibitemShut {NoStop}%
\bibitem [{\citenamefont {Yang}\ \emph {et~al.}(2015)\citenamefont {Yang},
  \citenamefont {Tian},\ and\ \citenamefont {Ji}}]{yang2015}%
  \BibitemOpen
  \bibfield  {author} {\bibinfo {author} {\bibfnamefont {D.}~\bibnamefont
  {Yang}}, \bibinfo {author} {\bibfnamefont {H.}~\bibnamefont {Tian}},\ and\
  \bibinfo {author} {\bibfnamefont {Y.}~\bibnamefont {Ji}},\ }\bibfield
  {title} {\bibinfo {title} {High-{{Q}} and high-sensitivity width-modulated
  photonic crystal single nanobeam air-mode cavity for refractive index
  sensing},\ }\href {https://doi.org/10.1364/AO.54.000001} {\bibfield
  {journal} {\bibinfo  {journal} {Appl. Opt., AO}\ }\textbf {\bibinfo {volume}
  {54}},\ \bibinfo {pages} {1} (\bibinfo {year} {2015})}\BibitemShut {NoStop}%
\bibitem [{\citenamefont {H{\o}j}\ \emph {et~al.}(2021)\citenamefont {H{\o}j},
  \citenamefont {Wang}, \citenamefont {Gao}, \citenamefont {Hoff},
  \citenamefont {Sigmund},\ and\ \citenamefont {Andersen}}]{hoj2021}%
  \BibitemOpen
  \bibfield  {author} {\bibinfo {author} {\bibfnamefont {D.}~\bibnamefont
  {H{\o}j}}, \bibinfo {author} {\bibfnamefont {F.}~\bibnamefont {Wang}},
  \bibinfo {author} {\bibfnamefont {W.}~\bibnamefont {Gao}}, \bibinfo {author}
  {\bibfnamefont {U.~B.}\ \bibnamefont {Hoff}}, \bibinfo {author}
  {\bibfnamefont {O.}~\bibnamefont {Sigmund}},\ and\ \bibinfo {author}
  {\bibfnamefont {U.~L.}\ \bibnamefont {Andersen}},\ }\bibfield  {title}
  {\bibinfo {title} {Ultra-coherent nanomechanical resonators based on inverse
  design},\ }\href@noop {} {\bibfield  {journal} {\bibinfo  {journal} {Nature
  communications}\ }\textbf {\bibinfo {volume} {12}},\ \bibinfo {pages} {5766}
  (\bibinfo {year} {2021})}\BibitemShut {NoStop}%
\bibitem [{\citenamefont {Tsaturyan}\ \emph {et~al.}(2017)\citenamefont
  {Tsaturyan}, \citenamefont {Barg}, \citenamefont {Polzik},\ and\
  \citenamefont {Schliesser}}]{tsaturyan2017}%
  \BibitemOpen
  \bibfield  {author} {\bibinfo {author} {\bibfnamefont {Y.}~\bibnamefont
  {Tsaturyan}}, \bibinfo {author} {\bibfnamefont {A.}~\bibnamefont {Barg}},
  \bibinfo {author} {\bibfnamefont {E.~S.}\ \bibnamefont {Polzik}},\ and\
  \bibinfo {author} {\bibfnamefont {A.}~\bibnamefont {Schliesser}},\ }\bibfield
   {title} {\bibinfo {title} {Ultracoherent nanomechanical resonators via soft
  clamping and dissipation dilution},\ }\href@noop {} {\bibfield  {journal}
  {\bibinfo  {journal} {Nature nanotechnology}\ }\textbf {\bibinfo {volume}
  {12}},\ \bibinfo {pages} {776} (\bibinfo {year} {2017})}\BibitemShut
  {NoStop}%
\bibitem [{\citenamefont {Engelsen}\ \emph {et~al.}(2024)\citenamefont
  {Engelsen}, \citenamefont {Beccari},\ and\ \citenamefont
  {Kippenberg}}]{engelsen2024}%
  \BibitemOpen
  \bibfield  {author} {\bibinfo {author} {\bibfnamefont {N.~J.}\ \bibnamefont
  {Engelsen}}, \bibinfo {author} {\bibfnamefont {A.}~\bibnamefont {Beccari}},\
  and\ \bibinfo {author} {\bibfnamefont {T.~J.}\ \bibnamefont {Kippenberg}},\
  }\bibfield  {title} {\bibinfo {title} {Ultrahigh-quality-factor micro-and
  nanomechanical resonators using dissipation dilution},\ }\href@noop {}
  {\bibfield  {journal} {\bibinfo  {journal} {Nature Nanotechnology}\ ,\
  \bibinfo {pages} {1}} (\bibinfo {year} {2024})}\BibitemShut {NoStop}%
\bibitem [{\citenamefont {Cupertino}\ \emph {et~al.}(2024)\citenamefont
  {Cupertino}, \citenamefont {Shin}, \citenamefont {Guo}, \citenamefont
  {Steeneken}, \citenamefont {Bessa},\ and\ \citenamefont
  {Norte}}]{cupertino2024}%
  \BibitemOpen
  \bibfield  {author} {\bibinfo {author} {\bibfnamefont {A.}~\bibnamefont
  {Cupertino}}, \bibinfo {author} {\bibfnamefont {D.}~\bibnamefont {Shin}},
  \bibinfo {author} {\bibfnamefont {L.}~\bibnamefont {Guo}}, \bibinfo {author}
  {\bibfnamefont {P.~G.}\ \bibnamefont {Steeneken}}, \bibinfo {author}
  {\bibfnamefont {M.~A.}\ \bibnamefont {Bessa}},\ and\ \bibinfo {author}
  {\bibfnamefont {R.~A.}\ \bibnamefont {Norte}},\ }\bibfield  {title} {\bibinfo
  {title} {Centimeter-scale nanomechanical resonators with low dissipation},\
  }\href@noop {} {\bibfield  {journal} {\bibinfo  {journal} {Nature
  Communications}\ }\textbf {\bibinfo {volume} {15}},\ \bibinfo {pages} {4255}
  (\bibinfo {year} {2024})}\BibitemShut {NoStop}%
\bibitem [{\citenamefont {Zaki}\ and\ \citenamefont
  {Basyooni}(2022)}]{zaki2022}%
  \BibitemOpen
  \bibfield  {author} {\bibinfo {author} {\bibfnamefont {S.~E.}\ \bibnamefont
  {Zaki}}\ and\ \bibinfo {author} {\bibfnamefont {M.~A.}\ \bibnamefont
  {Basyooni}},\ }\bibfield  {title} {\bibinfo {title} {Ultra-sensitive gas
  sensor based fano resonance modes in periodic and fibonacci quasi-periodic
  {{Pt}}/{{PtS2}} structures},\ }\href
  {https://doi.org/10.1038/s41598-022-13898-4} {\bibfield  {journal} {\bibinfo
  {journal} {Sci Rep}\ }\textbf {\bibinfo {volume} {12}},\ \bibinfo {pages}
  {9759} (\bibinfo {year} {2022})}\BibitemShut {NoStop}%
\bibitem [{\citenamefont {Sherif}\ and\ \citenamefont
  {Swillam}(2023)}]{sherif2023}%
  \BibitemOpen
  \bibfield  {author} {\bibinfo {author} {\bibfnamefont {S.~M.}\ \bibnamefont
  {Sherif}}\ and\ \bibinfo {author} {\bibfnamefont {M.~A.}\ \bibnamefont
  {Swillam}},\ }\bibfield  {title} {\bibinfo {title} {Silicon-based mid
  infrared on-chip gas sensor using {{Fano}} resonance of coupled plasmonic
  microcavities},\ }\href {https://doi.org/10.1038/s41598-023-38926-9}
  {\bibfield  {journal} {\bibinfo  {journal} {Sci Rep}\ }\textbf {\bibinfo
  {volume} {13}},\ \bibinfo {pages} {12311} (\bibinfo {year}
  {2023})}\BibitemShut {NoStop}%
\bibitem [{\citenamefont {Zaki}\ \emph {et~al.}(2020)\citenamefont {Zaki},
  \citenamefont {Mehaney}, \citenamefont {Hassanein},\ and\ \citenamefont
  {Aly}}]{zaki2020}%
  \BibitemOpen
  \bibfield  {author} {\bibinfo {author} {\bibfnamefont {S.~E.}\ \bibnamefont
  {Zaki}}, \bibinfo {author} {\bibfnamefont {A.}~\bibnamefont {Mehaney}},
  \bibinfo {author} {\bibfnamefont {H.~M.}\ \bibnamefont {Hassanein}},\ and\
  \bibinfo {author} {\bibfnamefont {A.~H.}\ \bibnamefont {Aly}},\ }\bibfield
  {title} {\bibinfo {title} {Fano resonance based defected {{1D}} phononic
  crystal for highly sensitive gas sensing applications},\ }\href
  {https://doi.org/10.1038/s41598-020-75076-8} {\bibfield  {journal} {\bibinfo
  {journal} {Sci Rep}\ }\textbf {\bibinfo {volume} {10}},\ \bibinfo {pages}
  {17979} (\bibinfo {year} {2020})}\BibitemShut {NoStop}%
\bibitem [{\citenamefont {Zhu}\ and\ \citenamefont {Wu}(2021)}]{zhu2021}%
  \BibitemOpen
  \bibfield  {author} {\bibinfo {author} {\bibfnamefont {J.}~\bibnamefont
  {Zhu}}\ and\ \bibinfo {author} {\bibfnamefont {C.}~\bibnamefont {Wu}},\
  }\bibfield  {title} {\bibinfo {title} {Optical refractive index sensor with
  {{Fano}} resonance based on original {{MIM}} waveguide structure},\ }\href
  {https://doi.org/10.1016/j.rinp.2021.103858} {\bibfield  {journal} {\bibinfo
  {journal} {Results in Physics}\ }\textbf {\bibinfo {volume} {21}},\ \bibinfo
  {pages} {103858} (\bibinfo {year} {2021})}\BibitemShut {NoStop}%
\bibitem [{\citenamefont {Tang}\ \emph {et~al.}(2017)\citenamefont {Tang},
  \citenamefont {Zhang}, \citenamefont {Wang}, \citenamefont {Hai},
  \citenamefont {Xue}, \citenamefont {Zhang},\ and\ \citenamefont
  {Yan}}]{tang2017}%
  \BibitemOpen
  \bibfield  {author} {\bibinfo {author} {\bibfnamefont {Y.}~\bibnamefont
  {Tang}}, \bibinfo {author} {\bibfnamefont {Z.}~\bibnamefont {Zhang}},
  \bibinfo {author} {\bibfnamefont {R.}~\bibnamefont {Wang}}, \bibinfo {author}
  {\bibfnamefont {Z.}~\bibnamefont {Hai}}, \bibinfo {author} {\bibfnamefont
  {C.}~\bibnamefont {Xue}}, \bibinfo {author} {\bibfnamefont {W.}~\bibnamefont
  {Zhang}},\ and\ \bibinfo {author} {\bibfnamefont {S.}~\bibnamefont {Yan}},\
  }\bibfield  {title} {\bibinfo {title} {Refractive index sensor based on fano
  resonances in metal-insulator-metal waveguides coupled with resonators},\
  }\href@noop {} {\bibfield  {journal} {\bibinfo  {journal} {Sensors}\ }\textbf
  {\bibinfo {volume} {17}},\ \bibinfo {pages} {784} (\bibinfo {year}
  {2017})}\BibitemShut {NoStop}%
\bibitem [{\citenamefont {Kong}\ \emph {et~al.}(2017)\citenamefont {Kong},
  \citenamefont {Wei}, \citenamefont {Liu},\ and\ \citenamefont
  {Wang}}]{kong2017}%
  \BibitemOpen
  \bibfield  {author} {\bibinfo {author} {\bibfnamefont {Y.}~\bibnamefont
  {Kong}}, \bibinfo {author} {\bibfnamefont {Q.}~\bibnamefont {Wei}}, \bibinfo
  {author} {\bibfnamefont {C.}~\bibnamefont {Liu}},\ and\ \bibinfo {author}
  {\bibfnamefont {S.}~\bibnamefont {Wang}},\ }\bibfield  {title} {\bibinfo
  {title} {Nanoscale temperature sensor based on {{Fano}} resonance in
  metal--insulator--metal waveguide},\ }\href
  {https://doi.org/10.1016/j.optcom.2016.09.041} {\bibfield  {journal}
  {\bibinfo  {journal} {Optics Communications}\ }\textbf {\bibinfo {volume}
  {384}},\ \bibinfo {pages} {85} (\bibinfo {year} {2017})}\BibitemShut
  {NoStop}%
\bibitem [{\citenamefont {Limonov}\ \emph {et~al.}(2017)\citenamefont
  {Limonov}, \citenamefont {Rybin}, \citenamefont {Poddubny},\ and\
  \citenamefont {Kivshar}}]{limonov2017}%
  \BibitemOpen
  \bibfield  {author} {\bibinfo {author} {\bibfnamefont {M.~F.}\ \bibnamefont
  {Limonov}}, \bibinfo {author} {\bibfnamefont {M.~V.}\ \bibnamefont {Rybin}},
  \bibinfo {author} {\bibfnamefont {A.~N.}\ \bibnamefont {Poddubny}},\ and\
  \bibinfo {author} {\bibfnamefont {Y.~S.}\ \bibnamefont {Kivshar}},\
  }\bibfield  {title} {\bibinfo {title} {Fano resonances in photonics},\
  }\href@noop {} {\bibfield  {journal} {\bibinfo  {journal} {Nature photonics}\
  }\textbf {\bibinfo {volume} {11}},\ \bibinfo {pages} {543} (\bibinfo {year}
  {2017})}\BibitemShut {NoStop}%
\bibitem [{\citenamefont {Yang}\ \emph {et~al.}(2024)\citenamefont {Yang},
  \citenamefont {Rajasekar},\ and\ \citenamefont {Sanju{\'a}n}}]{yang2024}%
  \BibitemOpen
  \bibfield  {author} {\bibinfo {author} {\bibfnamefont {J.}~\bibnamefont
  {Yang}}, \bibinfo {author} {\bibfnamefont {S.}~\bibnamefont {Rajasekar}},\
  and\ \bibinfo {author} {\bibfnamefont {M.~A.}\ \bibnamefont {Sanju{\'a}n}},\
  }\bibfield  {title} {\bibinfo {title} {Vibrational resonance: A review},\
  }\href@noop {} {\bibfield  {journal} {\bibinfo  {journal} {Physics Reports}\
  }\textbf {\bibinfo {volume} {1067}},\ \bibinfo {pages} {1} (\bibinfo {year}
  {2024})}\BibitemShut {NoStop}%
\bibitem [{\citenamefont {Pham}\ \emph {et~al.}(2020)\citenamefont {Pham},
  \citenamefont {Vaidyanathan},\ and\ \citenamefont {Kapitaniak}}]{pham2020}%
  \BibitemOpen
  \bibfield  {author} {\bibinfo {author} {\bibfnamefont {V.-T.}\ \bibnamefont
  {Pham}}, \bibinfo {author} {\bibfnamefont {S.}~\bibnamefont {Vaidyanathan}},\
  and\ \bibinfo {author} {\bibfnamefont {T.}~\bibnamefont {Kapitaniak}},\
  }\bibfield  {title} {\bibinfo {title} {Complexity, dynamics, control, and
  applications of nonlinear systems with multistability},\ }\href@noop {}
  {\bibfield  {journal} {\bibinfo  {journal} {Complexity}\ }\textbf {\bibinfo
  {volume} {2020}} (\bibinfo {year} {2020})}\BibitemShut {NoStop}%
\bibitem [{\citenamefont {Esmaeil~Zadeh}\ \emph {et~al.}(2021)\citenamefont
  {Esmaeil~Zadeh}, \citenamefont {Chang}, \citenamefont {Los}, \citenamefont
  {Gyger}, \citenamefont {Elshaari}, \citenamefont {Steinhauer}, \citenamefont
  {Dorenbos},\ and\ \citenamefont {Zwiller}}]{esmaeil2021}%
  \BibitemOpen
  \bibfield  {author} {\bibinfo {author} {\bibfnamefont {I.}~\bibnamefont
  {Esmaeil~Zadeh}}, \bibinfo {author} {\bibfnamefont {J.}~\bibnamefont
  {Chang}}, \bibinfo {author} {\bibfnamefont {J.~W.}\ \bibnamefont {Los}},
  \bibinfo {author} {\bibfnamefont {S.}~\bibnamefont {Gyger}}, \bibinfo
  {author} {\bibfnamefont {A.~W.}\ \bibnamefont {Elshaari}}, \bibinfo {author}
  {\bibfnamefont {S.}~\bibnamefont {Steinhauer}}, \bibinfo {author}
  {\bibfnamefont {S.~N.}\ \bibnamefont {Dorenbos}},\ and\ \bibinfo {author}
  {\bibfnamefont {V.}~\bibnamefont {Zwiller}},\ }\bibfield  {title} {\bibinfo
  {title} {Superconducting nanowire single-photon detectors: A perspective on
  evolution, state-of-the-art, future developments, and applications},\
  }\href@noop {} {\bibfield  {journal} {\bibinfo  {journal} {Applied Physics
  Letters}\ }\textbf {\bibinfo {volume} {118}} (\bibinfo {year}
  {2021})}\BibitemShut {NoStop}%
\bibitem [{\citenamefont {Pfenning}\ \emph {et~al.}(2022)\citenamefont
  {Pfenning}, \citenamefont {Kr{\"u}ger}, \citenamefont {Jabeen}, \citenamefont
  {Worschech}, \citenamefont {Hartmann},\ and\ \citenamefont
  {H{\"o}fling}}]{pfenning2022}%
  \BibitemOpen
  \bibfield  {author} {\bibinfo {author} {\bibfnamefont {A.}~\bibnamefont
  {Pfenning}}, \bibinfo {author} {\bibfnamefont {S.}~\bibnamefont
  {Kr{\"u}ger}}, \bibinfo {author} {\bibfnamefont {F.}~\bibnamefont {Jabeen}},
  \bibinfo {author} {\bibfnamefont {L.}~\bibnamefont {Worschech}}, \bibinfo
  {author} {\bibfnamefont {F.}~\bibnamefont {Hartmann}},\ and\ \bibinfo
  {author} {\bibfnamefont {S.}~\bibnamefont {H{\"o}fling}},\ }\bibfield
  {title} {\bibinfo {title} {Single-photon counting with semiconductor resonant
  tunneling devices},\ }\href@noop {} {\bibfield  {journal} {\bibinfo
  {journal} {Nanomaterials}\ }\textbf {\bibinfo {volume} {12}},\ \bibinfo
  {pages} {2358} (\bibinfo {year} {2022})}\BibitemShut {NoStop}%
\bibitem [{\citenamefont {Blakesley}\ \emph {et~al.}(2005)\citenamefont
  {Blakesley}, \citenamefont {See}, \citenamefont {Shields}, \citenamefont
  {Kardyna{\l}}, \citenamefont {Atkinson}, \citenamefont {Farrer},\ and\
  \citenamefont {Ritchie}}]{blakesley2005}%
  \BibitemOpen
  \bibfield  {author} {\bibinfo {author} {\bibfnamefont {J.}~\bibnamefont
  {Blakesley}}, \bibinfo {author} {\bibfnamefont {P.}~\bibnamefont {See}},
  \bibinfo {author} {\bibfnamefont {A.}~\bibnamefont {Shields}}, \bibinfo
  {author} {\bibfnamefont {B.}~\bibnamefont {Kardyna{\l}}}, \bibinfo {author}
  {\bibfnamefont {P.}~\bibnamefont {Atkinson}}, \bibinfo {author}
  {\bibfnamefont {I.}~\bibnamefont {Farrer}},\ and\ \bibinfo {author}
  {\bibfnamefont {D.}~\bibnamefont {Ritchie}},\ }\bibfield  {title} {\bibinfo
  {title} {Efficient single photon detection by quantum dot resonant tunneling
  diodes},\ }\href@noop {} {\bibfield  {journal} {\bibinfo  {journal} {Physical
  review letters}\ }\textbf {\bibinfo {volume} {94}},\ \bibinfo {pages}
  {067401} (\bibinfo {year} {2005})}\BibitemShut {NoStop}%
\bibitem [{\citenamefont {Dai}\ \emph {et~al.}(2009)\citenamefont {Dai},
  \citenamefont {Eom},\ and\ \citenamefont {Kim}}]{dai2009}%
  \BibitemOpen
  \bibfield  {author} {\bibinfo {author} {\bibfnamefont {M.~D.}\ \bibnamefont
  {Dai}}, \bibinfo {author} {\bibfnamefont {K.}~\bibnamefont {Eom}},\ and\
  \bibinfo {author} {\bibfnamefont {C.-W.}\ \bibnamefont {Kim}},\ }\bibfield
  {title} {\bibinfo {title} {Nanomechanical mass detection using nonlinear
  oscillations},\ }\href {https://doi.org/10.1063/1.3265731} {\bibfield
  {journal} {\bibinfo  {journal} {Applied Physics Letters}\ }\textbf {\bibinfo
  {volume} {95}},\ \bibinfo {pages} {203104} (\bibinfo {year}
  {2009})}\BibitemShut {NoStop}%
\bibitem [{\citenamefont {Dai}\ \emph {et~al.}(2012)\citenamefont {Dai},
  \citenamefont {Kim},\ and\ \citenamefont {Eom}}]{dai2012}%
  \BibitemOpen
  \bibfield  {author} {\bibinfo {author} {\bibfnamefont {M.~D.}\ \bibnamefont
  {Dai}}, \bibinfo {author} {\bibfnamefont {C.-W.}\ \bibnamefont {Kim}},\ and\
  \bibinfo {author} {\bibfnamefont {K.}~\bibnamefont {Eom}},\ }\bibfield
  {title} {\bibinfo {title} {Nonlinear vibration behavior of graphene
  resonators and their applications in sensitive mass detection},\ }\href@noop
  {} {\bibfield  {journal} {\bibinfo  {journal} {Nanoscale research letters}\
  }\textbf {\bibinfo {volume} {7}},\ \bibinfo {pages} {1} (\bibinfo {year}
  {2012})}\BibitemShut {NoStop}%
\bibitem [{\citenamefont {Kumar}\ \emph {et~al.}(2012)\citenamefont {Kumar},
  \citenamefont {Yang}, \citenamefont {Boley}, \citenamefont {Chiu},\ and\
  \citenamefont {Rhoads}}]{kumar2012}%
  \BibitemOpen
  \bibfield  {author} {\bibinfo {author} {\bibfnamefont {V.}~\bibnamefont
  {Kumar}}, \bibinfo {author} {\bibfnamefont {Y.}~\bibnamefont {Yang}},
  \bibinfo {author} {\bibfnamefont {J.~W.}\ \bibnamefont {Boley}}, \bibinfo
  {author} {\bibfnamefont {G.~T.-C.}\ \bibnamefont {Chiu}},\ and\ \bibinfo
  {author} {\bibfnamefont {J.~F.}\ \bibnamefont {Rhoads}},\ }\bibfield  {title}
  {\bibinfo {title} {Modeling, analysis, and experimental validation of a
  bifurcation-based microsensor},\ }\href@noop {} {\bibfield  {journal}
  {\bibinfo  {journal} {Journal of microelectromechanical systems}\ }\textbf
  {\bibinfo {volume} {21}},\ \bibinfo {pages} {549} (\bibinfo {year}
  {2012})}\BibitemShut {NoStop}%
\bibitem [{\citenamefont {Yuksel}\ \emph {et~al.}(2019)\citenamefont {Yuksel},
  \citenamefont {Orhan}, \citenamefont {Yanik}, \citenamefont {Ari},
  \citenamefont {Demir},\ and\ \citenamefont {Hanay}}]{yuksel2019}%
  \BibitemOpen
  \bibfield  {author} {\bibinfo {author} {\bibfnamefont {M.}~\bibnamefont
  {Yuksel}}, \bibinfo {author} {\bibfnamefont {E.}~\bibnamefont {Orhan}},
  \bibinfo {author} {\bibfnamefont {C.}~\bibnamefont {Yanik}}, \bibinfo
  {author} {\bibfnamefont {A.~B.}\ \bibnamefont {Ari}}, \bibinfo {author}
  {\bibfnamefont {A.}~\bibnamefont {Demir}},\ and\ \bibinfo {author}
  {\bibfnamefont {M.~S.}\ \bibnamefont {Hanay}},\ }\bibfield  {title} {\bibinfo
  {title} {Nonlinear nanomechanical mass spectrometry at the
  single-nanoparticle level},\ }\href@noop {} {\bibfield  {journal} {\bibinfo
  {journal} {Nano letters}\ }\textbf {\bibinfo {volume} {19}},\ \bibinfo
  {pages} {3583} (\bibinfo {year} {2019})}\BibitemShut {NoStop}%
\bibitem [{\citenamefont {{Al-Ghamdi}}\ \emph {et~al.}(2018)\citenamefont
  {{Al-Ghamdi}}, \citenamefont {Khater}, \citenamefont {Stewart}, \citenamefont
  {Alneamy}, \citenamefont {{Abdel-Rahman}},\ and\ \citenamefont
  {Penlidis}}]{al-ghamdi2018}%
  \BibitemOpen
  \bibfield  {author} {\bibinfo {author} {\bibfnamefont {M.~S.}\ \bibnamefont
  {{Al-Ghamdi}}}, \bibinfo {author} {\bibfnamefont {M.~E.}\ \bibnamefont
  {Khater}}, \bibinfo {author} {\bibfnamefont {K.~M.~E.}\ \bibnamefont
  {Stewart}}, \bibinfo {author} {\bibfnamefont {A.}~\bibnamefont {Alneamy}},
  \bibinfo {author} {\bibfnamefont {E.~M.}\ \bibnamefont {{Abdel-Rahman}}},\
  and\ \bibinfo {author} {\bibfnamefont {A.}~\bibnamefont {Penlidis}},\
  }\bibfield  {title} {\bibinfo {title} {Dynamic bifurcation {{MEMS}} gas
  sensors},\ }\href {https://doi.org/10.1088/1361-6439/aaedf9} {\bibfield
  {journal} {\bibinfo  {journal} {J. Micromech. Microeng.}\ }\textbf {\bibinfo
  {volume} {29}},\ \bibinfo {pages} {015005} (\bibinfo {year}
  {2018})}\BibitemShut {NoStop}%
\bibitem [{\citenamefont {Shama}\ \emph {et~al.}(2024)\citenamefont {Shama},
  \citenamefont {Rahmanian}, \citenamefont {Mouharrar}, \citenamefont
  {Abdelrahman}, \citenamefont {Elhady},\ and\ \citenamefont
  {Abdel-Rahman}}]{shama2024}%
  \BibitemOpen
  \bibfield  {author} {\bibinfo {author} {\bibfnamefont {Y.~S.}\ \bibnamefont
  {Shama}}, \bibinfo {author} {\bibfnamefont {S.}~\bibnamefont {Rahmanian}},
  \bibinfo {author} {\bibfnamefont {H.}~\bibnamefont {Mouharrar}}, \bibinfo
  {author} {\bibfnamefont {R.}~\bibnamefont {Abdelrahman}}, \bibinfo {author}
  {\bibfnamefont {A.}~\bibnamefont {Elhady}},\ and\ \bibinfo {author}
  {\bibfnamefont {E.~M.}\ \bibnamefont {Abdel-Rahman}},\ }\bibfield  {title}
  {\bibinfo {title} {Unraveling the nature of sensing in electrostatic mems gas
  sensors},\ }\href@noop {} {\bibfield  {journal} {\bibinfo  {journal}
  {Microsystems \& Nanoengineering}\ }\textbf {\bibinfo {volume} {10}},\
  \bibinfo {pages} {56} (\bibinfo {year} {2024})}\BibitemShut {NoStop}%
\bibitem [{\citenamefont {Zhang}\ \emph
  {et~al.}(2024{\natexlab{a}})\citenamefont {Zhang}, \citenamefont {Dong},
  \citenamefont {Wang},\ and\ \citenamefont {Huang}}]{zhang2024}%
  \BibitemOpen
  \bibfield  {author} {\bibinfo {author} {\bibfnamefont {M.-N.}\ \bibnamefont
  {Zhang}}, \bibinfo {author} {\bibfnamefont {L.}~\bibnamefont {Dong}},
  \bibinfo {author} {\bibfnamefont {L.-F.}\ \bibnamefont {Wang}},\ and\
  \bibinfo {author} {\bibfnamefont {Q.-A.}\ \bibnamefont {Huang}},\ }\bibfield
  {title} {\bibinfo {title} {Exceptional points enhance sensing in silicon
  micromechanical resonators},\ }\href@noop {} {\bibfield  {journal} {\bibinfo
  {journal} {Microsystems \& Nanoengineering}\ }\textbf {\bibinfo {volume}
  {10}},\ \bibinfo {pages} {12} (\bibinfo {year}
  {2024}{\natexlab{a}})}\BibitemShut {NoStop}%
\bibitem [{\citenamefont {Blaikie}\ \emph {et~al.}(2019)\citenamefont
  {Blaikie}, \citenamefont {Miller},\ and\ \citenamefont
  {Alem{\'a}n}}]{blaikie2019}%
  \BibitemOpen
  \bibfield  {author} {\bibinfo {author} {\bibfnamefont {A.}~\bibnamefont
  {Blaikie}}, \bibinfo {author} {\bibfnamefont {D.}~\bibnamefont {Miller}},\
  and\ \bibinfo {author} {\bibfnamefont {B.~J.}\ \bibnamefont {Alem{\'a}n}},\
  }\bibfield  {title} {\bibinfo {title} {A fast and sensitive room-temperature
  graphene nanomechanical bolometer},\ }\href@noop {} {\bibfield  {journal}
  {\bibinfo  {journal} {Nature communications}\ }\textbf {\bibinfo {volume}
  {10}},\ \bibinfo {pages} {4726} (\bibinfo {year} {2019})}\BibitemShut
  {NoStop}%
\bibitem [{\citenamefont {Zhang}\ \emph {et~al.}(2019)\citenamefont {Zhang},
  \citenamefont {Hosono}, \citenamefont {Nagai},\ and\ \citenamefont {Song
  S.-H.}}]{zhang2019}%
  \BibitemOpen
  \bibfield  {author} {\bibinfo {author} {\bibfnamefont {Y.}~\bibnamefont
  {Zhang}}, \bibinfo {author} {\bibfnamefont {S.}~\bibnamefont {Hosono}},
  \bibinfo {author} {\bibfnamefont {N.}~\bibnamefont {Nagai}},\ and\ \bibinfo
  {author} {\bibfnamefont {K.}~\bibnamefont {Song S.-H.}, \bibfnamefont
  {Hirakawa}},\ }\bibfield  {title} {\bibinfo {title} {Fast and sensitive
  bolometric terahertz detection at room temperature through thermomechanical
  transduction},\ }\href@noop {} {\bibfield  {journal} {\bibinfo  {journal} {J.
  Appl. Phys.}\ }\textbf {\bibinfo {volume} {125}},\ \bibinfo {pages} {151602}
  (\bibinfo {year} {2019})}\BibitemShut {NoStop}%
\bibitem [{\citenamefont {Vicarelli}\ \emph {et~al.}(2022)\citenamefont
  {Vicarelli}, \citenamefont {Tredicucci},\ and\ \citenamefont
  {Pitanti}}]{vicarelli2022}%
  \BibitemOpen
  \bibfield  {author} {\bibinfo {author} {\bibfnamefont {L.}~\bibnamefont
  {Vicarelli}}, \bibinfo {author} {\bibfnamefont {A.}~\bibnamefont
  {Tredicucci}},\ and\ \bibinfo {author} {\bibfnamefont {A.}~\bibnamefont
  {Pitanti}},\ }\bibfield  {title} {\bibinfo {title} {Micromechanical
  {{Bolometers}} for {{Subterahertz Detection}} at {{Room Temperature}}},\
  }\href {https://doi.org/10.1021/acsphotonics.1c01273} {\bibfield  {journal}
  {\bibinfo  {journal} {ACS Photonics}\ }\textbf {\bibinfo {volume} {9}},\
  \bibinfo {pages} {360} (\bibinfo {year} {2022})}\BibitemShut {NoStop}%
\bibitem [{\citenamefont {Piller}\ \emph {et~al.}(2022)\citenamefont {Piller},
  \citenamefont {Hiesberger}, \citenamefont {Wistrela}, \citenamefont
  {Martini}, \citenamefont {Luhmann},\ and\ \citenamefont
  {Schmid}}]{piller2022}%
  \BibitemOpen
  \bibfield  {author} {\bibinfo {author} {\bibfnamefont {M.}~\bibnamefont
  {Piller}}, \bibinfo {author} {\bibfnamefont {J.}~\bibnamefont {Hiesberger}},
  \bibinfo {author} {\bibfnamefont {E.}~\bibnamefont {Wistrela}}, \bibinfo
  {author} {\bibfnamefont {P.}~\bibnamefont {Martini}}, \bibinfo {author}
  {\bibfnamefont {N.}~\bibnamefont {Luhmann}},\ and\ \bibinfo {author}
  {\bibfnamefont {S.}~\bibnamefont {Schmid}},\ }\bibfield  {title} {\bibinfo
  {title} {Thermal ir detection with nanoelectromechanical silicon nitride
  trampoline resonators},\ }\href@noop {} {\bibfield  {journal} {\bibinfo
  {journal} {IEEE Sensors Journal}\ }\textbf {\bibinfo {volume} {23}},\
  \bibinfo {pages} {1066} (\bibinfo {year} {2022})}\BibitemShut {NoStop}%
\bibitem [{\citenamefont {Zhang}\ \emph
  {et~al.}(2024{\natexlab{b}})\citenamefont {Zhang}, \citenamefont
  {Yalavarthi}, \citenamefont {Giroux}, \citenamefont {Cui}, \citenamefont
  {Stephan}, \citenamefont {Maleki}, \citenamefont {Weck}, \citenamefont
  {M{\'e}nard},\ and\ \citenamefont {St-Gelais}}]{zhang2024b}%
  \BibitemOpen
  \bibfield  {author} {\bibinfo {author} {\bibfnamefont {C.}~\bibnamefont
  {Zhang}}, \bibinfo {author} {\bibfnamefont {E.~K.}\ \bibnamefont
  {Yalavarthi}}, \bibinfo {author} {\bibfnamefont {M.}~\bibnamefont {Giroux}},
  \bibinfo {author} {\bibfnamefont {W.}~\bibnamefont {Cui}}, \bibinfo {author}
  {\bibfnamefont {M.}~\bibnamefont {Stephan}}, \bibinfo {author} {\bibfnamefont
  {A.}~\bibnamefont {Maleki}}, \bibinfo {author} {\bibfnamefont
  {A.}~\bibnamefont {Weck}}, \bibinfo {author} {\bibfnamefont {J.-M.}\
  \bibnamefont {M{\'e}nard}},\ and\ \bibinfo {author} {\bibfnamefont
  {R.}~\bibnamefont {St-Gelais}},\ }\bibfield  {title} {\bibinfo {title} {High
  detectivity terahertz radiation sensing using frequency-noise-optimized
  nanomechanical resonators},\ }\href@noop {} {\bibfield  {journal} {\bibinfo
  {journal} {arXiv preprint arXiv:2401.16503}\ } (\bibinfo {year}
  {2024}{\natexlab{b}})}\BibitemShut {NoStop}%
\bibitem [{\citenamefont {Duraffourg}\ \emph {et~al.}(2018)\citenamefont
  {Duraffourg}, \citenamefont {Laurent}, \citenamefont {Moulet}, \citenamefont
  {Arcamone},\ and\ \citenamefont {Yon}}]{duraffourg2018}%
  \BibitemOpen
  \bibfield  {author} {\bibinfo {author} {\bibfnamefont {L.}~\bibnamefont
  {Duraffourg}}, \bibinfo {author} {\bibfnamefont {L.}~\bibnamefont {Laurent}},
  \bibinfo {author} {\bibfnamefont {J.-S.}\ \bibnamefont {Moulet}}, \bibinfo
  {author} {\bibfnamefont {J.}~\bibnamefont {Arcamone}},\ and\ \bibinfo
  {author} {\bibfnamefont {J.-J.}\ \bibnamefont {Yon}},\ }\bibfield  {title}
  {\bibinfo {title} {Array of resonant electromechanical nanosystems: A
  technological breakthrough for uncooled infrared imaging},\ }\href@noop {}
  {\bibfield  {journal} {\bibinfo  {journal} {Micromachines}\ }\textbf
  {\bibinfo {volume} {9}},\ \bibinfo {pages} {401} (\bibinfo {year}
  {2018})}\BibitemShut {NoStop}%
\bibitem [{\citenamefont {Gregorat}\ \emph {et~al.}(2024)\citenamefont
  {Gregorat}, \citenamefont {Cautero}, \citenamefont {Vicarelli}, \citenamefont
  {Giuressi}, \citenamefont {Bagolini}, \citenamefont {Tredicucci},
  \citenamefont {Cautero},\ and\ \citenamefont {Pitanti}}]{gregorat2024}%
  \BibitemOpen
  \bibfield  {author} {\bibinfo {author} {\bibfnamefont {L.}~\bibnamefont
  {Gregorat}}, \bibinfo {author} {\bibfnamefont {M.}~\bibnamefont {Cautero}},
  \bibinfo {author} {\bibfnamefont {L.}~\bibnamefont {Vicarelli}}, \bibinfo
  {author} {\bibfnamefont {D.}~\bibnamefont {Giuressi}}, \bibinfo {author}
  {\bibfnamefont {A.}~\bibnamefont {Bagolini}}, \bibinfo {author}
  {\bibfnamefont {A.}~\bibnamefont {Tredicucci}}, \bibinfo {author}
  {\bibfnamefont {G.}~\bibnamefont {Cautero}},\ and\ \bibinfo {author}
  {\bibfnamefont {A.}~\bibnamefont {Pitanti}},\ }\bibfield  {title} {\bibinfo
  {title} {Highly dispersive multiplexed micromechanical device array for
  spatially resolved sensing and actuation},\ }\href@noop {} {\bibfield
  {journal} {\bibinfo  {journal} {Microsyst. Nanoeng.}\ } (\bibinfo {year}
  {2024})}\BibitemShut {NoStop}%
\bibitem [{\citenamefont {Li}\ \emph {et~al.}(2023)\citenamefont {Li},
  \citenamefont {Zhang},\ and\ \citenamefont {Hirakawa}}]{li2023}%
  \BibitemOpen
  \bibfield  {author} {\bibinfo {author} {\bibfnamefont {C.}~\bibnamefont
  {Li}}, \bibinfo {author} {\bibfnamefont {Y.}~\bibnamefont {Zhang}},\ and\
  \bibinfo {author} {\bibfnamefont {K.}~\bibnamefont {Hirakawa}},\ }\bibfield
  {title} {\bibinfo {title} {Terahertz {{Detectors Using Microelectromechanical
  System Resonators}}},\ }\href {https://doi.org/10.3390/s23135938} {\bibfield
  {journal} {\bibinfo  {journal} {Sensors}\ }\textbf {\bibinfo {volume} {23}},\
  \bibinfo {pages} {5938} (\bibinfo {year} {2023})}\BibitemShut {NoStop}%
\bibitem [{\citenamefont {Schmid}\ \emph {et~al.}(2016)\citenamefont {Schmid},
  \citenamefont {Villanueva},\ and\ \citenamefont {Roukes}}]{schmid2016}%
  \BibitemOpen
  \bibfield  {author} {\bibinfo {author} {\bibfnamefont {S.}~\bibnamefont
  {Schmid}}, \bibinfo {author} {\bibfnamefont {L.~G.}\ \bibnamefont
  {Villanueva}},\ and\ \bibinfo {author} {\bibfnamefont {M.~L.}\ \bibnamefont
  {Roukes}},\ }\href@noop {} {\emph {\bibinfo {title} {Fundamentals of
  nanomechanical resonators}}},\ Vol.~\bibinfo {volume} {49}\ (\bibinfo
  {publisher} {Springer},\ \bibinfo {year} {2016})\BibitemShut {NoStop}%
\bibitem [{\citenamefont {Schuster}(2008)}]{Schuster2008}%
  \BibitemOpen
  \bibfield  {author} {\bibinfo {author} {\bibfnamefont {H.~G.}\ \bibnamefont
  {Schuster}},\ }\href@noop {} {\emph {\bibinfo {title} {Reviews of nonlinear
  dynamics and complexity}}}\ (\bibinfo  {publisher} {Wiley Online Library},\
  \bibinfo {year} {2008})\BibitemShut {NoStop}%
\bibitem [{\citenamefont {Kanellopulos}\ \emph {et~al.}(2024)\citenamefont
  {Kanellopulos}, \citenamefont {Ladinig}, \citenamefont {Emminger},
  \citenamefont {Martini}, \citenamefont {West},\ and\ \citenamefont
  {Schmid}}]{kanellopulos2024}%
  \BibitemOpen
  \bibfield  {author} {\bibinfo {author} {\bibfnamefont {K.}~\bibnamefont
  {Kanellopulos}}, \bibinfo {author} {\bibfnamefont {F.}~\bibnamefont
  {Ladinig}}, \bibinfo {author} {\bibfnamefont {S.}~\bibnamefont {Emminger}},
  \bibinfo {author} {\bibfnamefont {P.}~\bibnamefont {Martini}}, \bibinfo
  {author} {\bibfnamefont {R.~G.}\ \bibnamefont {West}},\ and\ \bibinfo
  {author} {\bibfnamefont {S.}~\bibnamefont {Schmid}},\ }\bibfield  {title}
  {\bibinfo {title} {Comparative analysis of nanomechanical resonators:
  Sensitivity, response time, and practical considerations in photothermal
  sensing},\ }\href@noop {} {\bibfield  {journal} {\bibinfo  {journal} {arXiv
  preprint arXiv:2406.03295}\ } (\bibinfo {year} {2024})}\BibitemShut {NoStop}%
\bibitem [{\citenamefont {Norte}\ \emph {et~al.}(2016)\citenamefont {Norte},
  \citenamefont {Moura},\ and\ \citenamefont {Gr{\"o}blacher}}]{norte2016}%
  \BibitemOpen
  \bibfield  {author} {\bibinfo {author} {\bibfnamefont {R.~A.}\ \bibnamefont
  {Norte}}, \bibinfo {author} {\bibfnamefont {J.~P.}\ \bibnamefont {Moura}},\
  and\ \bibinfo {author} {\bibfnamefont {S.}~\bibnamefont {Gr{\"o}blacher}},\
  }\bibfield  {title} {\bibinfo {title} {Mechanical resonators for quantum
  optomechanics experiments at room temperature},\ }\href@noop {} {\bibfield
  {journal} {\bibinfo  {journal} {Physical review letters}\ }\textbf {\bibinfo
  {volume} {116}},\ \bibinfo {pages} {147202} (\bibinfo {year}
  {2016})}\BibitemShut {NoStop}%
\bibitem [{\citenamefont {Reinhardt}\ \emph {et~al.}(2016)\citenamefont
  {Reinhardt}, \citenamefont {M{\"u}ller}, \citenamefont {Bourassa},\ and\
  \citenamefont {Sankey}}]{reinhardt2016}%
  \BibitemOpen
  \bibfield  {author} {\bibinfo {author} {\bibfnamefont {C.}~\bibnamefont
  {Reinhardt}}, \bibinfo {author} {\bibfnamefont {T.}~\bibnamefont
  {M{\"u}ller}}, \bibinfo {author} {\bibfnamefont {A.}~\bibnamefont
  {Bourassa}},\ and\ \bibinfo {author} {\bibfnamefont {J.~C.}\ \bibnamefont
  {Sankey}},\ }\bibfield  {title} {\bibinfo {title} {Ultralow-noise sin
  trampoline resonators for sensing and optomechanics},\ }\href@noop {}
  {\bibfield  {journal} {\bibinfo  {journal} {Physical Review X}\ }\textbf
  {\bibinfo {volume} {6}},\ \bibinfo {pages} {021001} (\bibinfo {year}
  {2016})}\BibitemShut {NoStop}%
\bibitem [{\citenamefont {Venstra}\ \emph {et~al.}(2009)\citenamefont
  {Venstra}, \citenamefont {Westra}, \citenamefont {Gavan},\ and\ \citenamefont
  {Van Der~Zant}}]{venstra2009}%
  \BibitemOpen
  \bibfield  {author} {\bibinfo {author} {\bibfnamefont {W.}~\bibnamefont
  {Venstra}}, \bibinfo {author} {\bibfnamefont {H.}~\bibnamefont {Westra}},
  \bibinfo {author} {\bibfnamefont {K.~B.}\ \bibnamefont {Gavan}},\ and\
  \bibinfo {author} {\bibfnamefont {H.}~\bibnamefont {Van Der~Zant}},\
  }\bibfield  {title} {\bibinfo {title} {Magnetomotive drive and detection of
  clamped-clamped mechanical resonators in water},\ }\href@noop {} {\bibfield
  {journal} {\bibinfo  {journal} {Applied Physics Letters}\ }\textbf {\bibinfo
  {volume} {95}} (\bibinfo {year} {2009})}\BibitemShut {NoStop}%
\bibitem [{\citenamefont {Chien}\ \emph {et~al.}(2020)\citenamefont {Chien},
  \citenamefont {Steurer}, \citenamefont {Sadeghi}, \citenamefont {Cazier},\
  and\ \citenamefont {Schmid}}]{chien2020}%
  \BibitemOpen
  \bibfield  {author} {\bibinfo {author} {\bibfnamefont {M.-H.}\ \bibnamefont
  {Chien}}, \bibinfo {author} {\bibfnamefont {J.}~\bibnamefont {Steurer}},
  \bibinfo {author} {\bibfnamefont {P.}~\bibnamefont {Sadeghi}}, \bibinfo
  {author} {\bibfnamefont {N.}~\bibnamefont {Cazier}},\ and\ \bibinfo {author}
  {\bibfnamefont {S.}~\bibnamefont {Schmid}},\ }\bibfield  {title} {\bibinfo
  {title} {Nanoelectromechanical position-sensitive detector with picometer
  resolution},\ }\href@noop {} {\bibfield  {journal} {\bibinfo  {journal} {ACS
  photonics}\ }\textbf {\bibinfo {volume} {7}},\ \bibinfo {pages} {2197}
  (\bibinfo {year} {2020})}\BibitemShut {NoStop}%
\bibitem [{\citenamefont {Mura}\ \emph {et~al.}(2023)\citenamefont {Mura},
  \citenamefont {Lamberti}, \citenamefont {Tucci}, \citenamefont {Jorudas},
  \citenamefont {Cojocari}, \citenamefont {Fedorov},\ and\ \citenamefont
  {Kuzhir}}]{lamura2023}%
  \BibitemOpen
  \bibfield  {author} {\bibinfo {author} {\bibfnamefont {M.~L.}\ \bibnamefont
  {Mura}}, \bibinfo {author} {\bibfnamefont {P.}~\bibnamefont {Lamberti}},
  \bibinfo {author} {\bibfnamefont {V.}~\bibnamefont {Tucci}}, \bibinfo
  {author} {\bibfnamefont {J.}~\bibnamefont {Jorudas}}, \bibinfo {author}
  {\bibfnamefont {M.}~\bibnamefont {Cojocari}}, \bibinfo {author}
  {\bibfnamefont {G.}~\bibnamefont {Fedorov}},\ and\ \bibinfo {author}
  {\bibfnamefont {P.}~\bibnamefont {Kuzhir}},\ }\bibfield  {title} {\bibinfo
  {title} {Exploring the {{Impact}} of {{Absorber Material}} on the
  {{Performance}} of a {{Terahertz Microbolometer}} by {{Finite Element
  Analysis}}},\ }in\ \href {https://doi.org/10.1109/NMDC57951.2023.10343843}
  {\emph {\bibinfo {booktitle} {2023 {{IEEE Nanotechnology Materials}} and
  {{Devices Conference}} ({{NMDC}})}}}\ (\bibinfo {year} {2023})\ pp.\ \bibinfo
  {pages} {520--524}\BibitemShut {NoStop}%
\bibitem [{\citenamefont {Haddadi.~M}\ \emph {et~al.}(2022)\citenamefont
  {Haddadi.~M}, \citenamefont {Das}, \citenamefont {Jeong}, \citenamefont
  {Kim},\ and\ \citenamefont {Kim}}]{Haddadi2022}%
  \BibitemOpen
  \bibfield  {author} {\bibinfo {author} {\bibfnamefont {M.}~\bibnamefont
  {Haddadi.~M}}, \bibinfo {author} {\bibfnamefont {B.}~\bibnamefont {Das}},
  \bibinfo {author} {\bibfnamefont {J.}~\bibnamefont {Jeong}}, \bibinfo
  {author} {\bibfnamefont {S.}~\bibnamefont {Kim}},\ and\ \bibinfo {author}
  {\bibfnamefont {D.-S.}\ \bibnamefont {Kim}},\ }\bibfield  {title} {\bibinfo
  {title} {Near-maximum microwave absorption in a thin metal film at the
  pseudo-free-standing limit},\ }\href@noop {} {\bibfield  {journal} {\bibinfo
  {journal} {Scientific Reports}\ }\textbf {\bibinfo {volume} {12}},\ \bibinfo
  {pages} {18386} (\bibinfo {year} {2022})}\BibitemShut {NoStop}%
\bibitem [{\citenamefont {Kaplas}\ and\ \citenamefont
  {Kuzhir}(2017)}]{kaplas2017}%
  \BibitemOpen
  \bibfield  {author} {\bibinfo {author} {\bibfnamefont {T.}~\bibnamefont
  {Kaplas}}\ and\ \bibinfo {author} {\bibfnamefont {P.}~\bibnamefont
  {Kuzhir}},\ }\bibfield  {title} {\bibinfo {title} {Ultra-{{Thin Pyrocarbon
  Films}} as a {{Versatile Coating Material}}},\ }\href
  {https://doi.org/10.1186/s11671-017-1896-0} {\bibfield  {journal} {\bibinfo
  {journal} {Nanoscale Res Lett}\ }\textbf {\bibinfo {volume} {12}},\ \bibinfo
  {pages} {121} (\bibinfo {year} {2017})}\BibitemShut {NoStop}%
\bibitem [{\citenamefont {Kaplas}\ and\ \citenamefont
  {Svirko}(2012)}]{kaplas2012}%
  \BibitemOpen
  \bibfield  {author} {\bibinfo {author} {\bibfnamefont {T.}~\bibnamefont
  {Kaplas}}\ and\ \bibinfo {author} {\bibfnamefont {Y.~P.}\ \bibnamefont
  {Svirko}},\ }\bibfield  {title} {\bibinfo {title} {Direct deposition of
  semitransparent conducting pyrolytic carbon films},\ }\href
  {https://doi.org/10.1117/1.JNP.6.061703} {\bibfield  {journal} {\bibinfo
  {journal} {JNP}\ }\textbf {\bibinfo {volume} {6}},\ \bibinfo {pages} {061703}
  (\bibinfo {year} {2012})}\BibitemShut {NoStop}%
\bibitem [{\citenamefont {Jorudas}\ \emph {et~al.}(2024)\citenamefont
  {Jorudas}, \citenamefont {Rehman}, \citenamefont {Cojocari}, \citenamefont
  {Pashnev}, \citenamefont {Urbanowicz}, \citenamefont {Ka{\v s}alynas},
  \citenamefont {Bertoni}, \citenamefont {Vicarelli}, \citenamefont {Pitanti},
  \citenamefont {Malykhin}, \citenamefont {Svirko}, \citenamefont {Kuzhir},\
  and\ \citenamefont {Fedorov}}]{jorudas2024}%
  \BibitemOpen
  \bibfield  {author} {\bibinfo {author} {\bibfnamefont {J.}~\bibnamefont
  {Jorudas}}, \bibinfo {author} {\bibfnamefont {H.}~\bibnamefont {Rehman}},
  \bibinfo {author} {\bibfnamefont {M.}~\bibnamefont {Cojocari}}, \bibinfo
  {author} {\bibfnamefont {D.}~\bibnamefont {Pashnev}}, \bibinfo {author}
  {\bibfnamefont {A.}~\bibnamefont {Urbanowicz}}, \bibinfo {author}
  {\bibfnamefont {I.}~\bibnamefont {Ka{\v s}alynas}}, \bibinfo {author}
  {\bibfnamefont {B.}~\bibnamefont {Bertoni}}, \bibinfo {author} {\bibfnamefont
  {L.}~\bibnamefont {Vicarelli}}, \bibinfo {author} {\bibfnamefont
  {A.}~\bibnamefont {Pitanti}}, \bibinfo {author} {\bibfnamefont
  {S.}~\bibnamefont {Malykhin}}, \bibinfo {author} {\bibfnamefont
  {Y.}~\bibnamefont {Svirko}}, \bibinfo {author} {\bibfnamefont
  {P.}~\bibnamefont {Kuzhir}},\ and\ \bibinfo {author} {\bibfnamefont
  {G.}~\bibnamefont {Fedorov}},\ }\bibfield  {title} {\bibinfo {title}
  {Ultra-broadband absorbance of nanometer-thin pyrolyzed-carbon film on
  silicon nitride membrane},\ }\href {https://doi.org/10.1088/1361-6528/ad4157}
  {\bibfield  {journal} {\bibinfo  {journal} {Nanotechnology}\ }\textbf
  {\bibinfo {volume} {35}},\ \bibinfo {pages} {305705} (\bibinfo {year}
  {2024})}\BibitemShut {NoStop}%
\bibitem [{\citenamefont {Martini}\ \emph {et~al.}(2025)\citenamefont
  {Martini}, \citenamefont {Kanellopulos}, \citenamefont {Emminger},
  \citenamefont {Luhmann}, \citenamefont {Piller}, \citenamefont {West},\ and\
  \citenamefont {Schmid}}]{Martini2025}%
  \BibitemOpen
  \bibfield  {author} {\bibinfo {author} {\bibfnamefont {P.}~\bibnamefont
  {Martini}}, \bibinfo {author} {\bibfnamefont {K.}~\bibnamefont
  {Kanellopulos}}, \bibinfo {author} {\bibfnamefont {S.}~\bibnamefont
  {Emminger}}, \bibinfo {author} {\bibfnamefont {N.}~\bibnamefont {Luhmann}},
  \bibinfo {author} {\bibfnamefont {M.}~\bibnamefont {Piller}}, \bibinfo
  {author} {\bibfnamefont {R.~G.}\ \bibnamefont {West}},\ and\ \bibinfo
  {author} {\bibfnamefont {S.}~\bibnamefont {Schmid}},\ }\bibfield  {title}
  {\bibinfo {title} {{Uncooled thermal infrared detection near the fundamental
  limit using a silicon nitride nanomechanical resonator with a broadband
  absorber}},\ }\href {https://doi.org/10.1038/s42005-025-02093-2} {\bibfield
  {journal} {\bibinfo  {journal} {Commun. Phys.}\ }\textbf {\bibinfo {volume}
  {8}},\ \bibinfo {pages} {1} (\bibinfo {year} {2025})}\BibitemShut {NoStop}%
\bibitem [{\citenamefont {Ordal}\ \emph {et~al.}(1983)\citenamefont {Ordal},
  \citenamefont {Long}, \citenamefont {Bell}, \citenamefont {Bell},
  \citenamefont {Bell}, \citenamefont {Alexander},\ and\ \citenamefont
  {Ward}}]{ordal1983}%
  \BibitemOpen
  \bibfield  {author} {\bibinfo {author} {\bibfnamefont {M.~A.}\ \bibnamefont
  {Ordal}}, \bibinfo {author} {\bibfnamefont {L.}~\bibnamefont {Long}},
  \bibinfo {author} {\bibfnamefont {R.~J.}\ \bibnamefont {Bell}}, \bibinfo
  {author} {\bibfnamefont {S.}~\bibnamefont {Bell}}, \bibinfo {author}
  {\bibfnamefont {R.}~\bibnamefont {Bell}}, \bibinfo {author} {\bibfnamefont
  {R.~W.}\ \bibnamefont {Alexander}},\ and\ \bibinfo {author} {\bibfnamefont
  {C.}~\bibnamefont {Ward}},\ }\bibfield  {title} {\bibinfo {title} {Optical
  properties of the metals al, co, cu, au, fe, pb, ni, pd, pt, ag, ti, and w in
  the infrared and far infrared},\ }\href@noop {} {\bibfield  {journal}
  {\bibinfo  {journal} {Applied optics}\ }\textbf {\bibinfo {volume} {22}},\
  \bibinfo {pages} {1099} (\bibinfo {year} {1983})}\BibitemShut {NoStop}%
\bibitem [{\citenamefont {Rogalski}(2010)}]{rogalski2010infrared}%
  \BibitemOpen
  \bibfield  {author} {\bibinfo {author} {\bibfnamefont {A.}~\bibnamefont
  {Rogalski}},\ }\href@noop {} {\emph {\bibinfo {title} {Infrared
  Detectors}}},\ \bibinfo {edition} {2nd}\ ed.\ (\bibinfo  {publisher} {CRC
  Press},\ \bibinfo {address} {Boca Raton, FL},\ \bibinfo {year}
  {2010})\BibitemShut {NoStop}%
\bibitem [{VDI()}]{VDI}%
  \BibitemOpen
  \href@noop {} {\bibinfo {title} {Virginia diodes detectors}},\ \bibinfo
  {howpublished}
  {\url{https://www.vadiodes.com/en/products/detectors}}\BibitemShut {NoStop}%
\end{thebibliography}%

\end{document}